\documentclass[journal]{IEEEtran}
\usepackage{amsmath,amsfonts}
\usepackage{algorithmic}
\usepackage{algorithm}
\usepackage{array}
\usepackage[caption=false,font=normalsize,labelfont=sf,textfont=sf]{subfig}
\usepackage{textcomp}
\usepackage{stfloats}
\usepackage{url}
\usepackage{verbatim}
\usepackage{graphicx}
\usepackage{caption}
\usepackage{cite}
\usepackage{color}
\usepackage{standalone} 
\usepackage{tikz}
\usepackage{soul}
\usepackage{xcolor}
\sethlcolor{yellow}

\usetikzlibrary{angles, quotes, calc}
\usepackage{pgfplots}
\usepgfplotslibrary{groupplots}
\usepgfplotslibrary{fillbetween}
\pgfplotsset{compat=1.18}
\usetikzlibrary{patterns}
\usetikzlibrary{fillbetween}
\usetikzlibrary{backgrounds}

\usetikzlibrary{decorations.markings}

\begin{document}

\title{A Model Predictive Approach for Enhancing Transient Stability of Grid-Forming Converters}

\author{Ali Arjomandi-Nezhad,~\IEEEmembership{Student Member,~IEEE,}
\and
Yifei Guo,~\IEEEmembership{Member,~IEEE,}
\and
Bikash C. Pal,~\IEEEmembership{Fellow,~IEEE,}
\and
Damiano Varagnolo,~\IEEEmembership{Member,~IEEE}
\thanks{This work was supported in part by the European Union’s Horizon 2020 Research and Innovation Programme under the Marie Skłodowska-Curie Grant 956433, in part by the Resilient Operation of Sustainable Energy Systems (ROSES) U.K.-China (EPSRC-NSFC) Programme on Sustainable Energy Supply under Grants EP/T021713/1 and NSFC-52061635102, and in part by the Royal Society under Grant RG\textbackslash R2\textbackslash 232398. For the purpose of open access, the authors have applied a Creative Commons Attribution (CC BY) license to any Accepted Manuscript version arising.
}}

\markboth{}
{Shell \MakeLowercase{\textit{et al.}}: A Sample Article Using IEEEtran.cls for IEEE Journals}


\maketitle

\begin{abstract}
A model predictive control (MPC) method for enhancing post-fault transient stability of a grid-forming (GFM) inverter based resources (IBRs)  is developed in this paper. This proposed controller is activated as soon as the converter enters into the post-fault current-saturation mode. It aims at mitigating the instability arising from insufficient deceleration due to current saturation and thus improving the transient stability of a GFM-IBR. The MPC approach optimises the post-fault trajectory of GFM IBRs by introducing appropriate corrective phase angle jumps and active power references where the post-fault dynamics of GFM IBRs are addressed. These two signals provide controllability over GFM IBR's post-fault trajectory. This paper addresses the mitigation of oscillations between current-saturation mode and normal mode by forced saturation if conditions for remaining in the normal mode do not hold. The performance of the proposal is tested via dynamic simulations under various grid conditions and compared with other existing strategies. The results demonstrate significant improvement in transient stability.
\end{abstract}

\begin{IEEEkeywords}
Current saturation, grid-forming (GFMs) converters, model predictive control, synchronization, transient stability.
\end{IEEEkeywords}

\section{Introduction}

In response to climate change concerns, many countries have outlined national strategies to achieve net-zero emissions in power systems \cite{UKNetZero, DenmarkNetZero}. This largely relies on decarbonizing energy supply by developing renewables and phasing out fossil fuels. Unlike traditional fuel-based synchronous generators (SGs) interfaced resources, most renewables (such as wind and solar) are interfaced with power grids through power electronics, termed inverter-based resources (IBRs).  Nearly all the IBRs in service today are ‘grid-following’ (GFL), which inherently do not contribute to power system inertia and strength \cite{GFMintro}. This poses major challenges for system stability as the generation mix continues to evolve towards lower shares of SGs, in the sense that power disturbances cause faster frequency variation due to the low system inertia, and  voltage disturbances are propagated much further due to the low grid strength. To maintain the stability of the electrical system, grid-forming (GFM) technologies, which allow IBRs to contribute to system stabilisation by providing virtual inertia, voltage, and frequency support, are recommended to be utilized \cite{Lasseter2020, Wang2023, Pan2023}.

 A major challenge arises from the limited over-current capabilities of inverters \cite{Kkuni} , which inhibits their ability to provide sufficient active and reactive power during and after large disturbances (e.g., grid faults). As a result, transient stability is adversely affected. Oversizing the converter to achieve higher fault currents is a very costly option. On the other hand, the transient behaviors of GFM IBRs are mainly dictated by their control strategies unlike SGs of which the behaviors are tightly linked to the rotating physical components. Thus, it is more promising and cost-effective to enhance transient stability by refining the control system design, which admits a larger degree of freedom and flexibility.

A substantial body of literature has been dedicated to the transient stability enhancement of GFM IBRs. References \cite{Luo2022, Chen2022,Huang2019, Si2023, Ge2023} proposed modifying the GFM IBR's control loops for transient stability enhancement. However, these approaches did not consider the impact of current saturation during and after close in fault on transient stability. In practical implementation of GFM IBRs, current saturation is an essence to protect power electronic switches from over-current. Neglecting this saturation in transient stability studies results in misleading analyses.
There are two typical categories of current saturation \cite{Quria2020}: (a) virtual impedance-based approaches, which emulate a Virtual Impedance (VI) after the reference terminal voltage to reduce the over-current, and (b) current reference saturation (CRS), which puts a hard limit on the current reference generated by the voltage controller of the GFM IBR. The authors of \cite{Xiong2021, Zhao2022, Li2022, Jin2022, Zou2023} proposed transient stability enhancement methods that consider the VI-based current saturation. While the VI-based methods might exhibit a higher margin of stability compared to current reference saturation (CRS)-based approaches, they may fail to promptly limit the inverter's output current. As a result, the inverter may be overloaded until VI limits the current \cite{Quria2020}.

It is shown in \cite{Rokrok2022} that if the angle of current saturation in CRS approach is not set correctly, GFM IBRs may be stuck in the post-fault current-saturation operation mode. For this purpose, reference \cite{Rokrok2022} proposed to optimise the current angle to prevent the GFM IBRs from being locked in the current-saturation mode. The authors in \cite{Sun2022} proposed to change the reference power of GFM IBRs again to enhance transient stability. The angle of current saturation was optimized in \cite{Liu2023} to maximize Lyapunov-assessed transient stability region of GFM IBRs. Data-driven predictive control was proposed in \cite{Shen2022}, where a corrective power reference change was introduced to achieve enhanced transient stability. The possible benefits of an optimal direct corrective change of the angle alongside power reference change was not considered in \cite{Rokrok2022, Sun2022, Liu2023, Shen2022}. Therefore, the whole potential of modifying the synchronization control for transient stability enhancement was not investigated in these studies.

In this paper, the goal is to untap this opportunity by modifying the post-fault trajectory of GFM IBRs via introducing appropriate corrective phase jumps and power reference change into the Active Power Control (APC) loop of GFM IBRs. These two corrective signals directly modify the APC angle and frequency, respectively. To avoid compromising with steady-state operation and small signal stability \cite{Baghaee2017, Jamali2022}, the corrective terms are activated only when GFM IBR is in its post-fault current-saturation mode. The values of these two signals are calculated by a Model Predictive Control (MPC) approach. The main contributions of the paper are as follows:

\begin{enumerate}
    \item Corrective phase jumps and power reference change are introduced to the APC when a GFM IBR is in its post-fault current-saturation mode. These signals are optimized through an MPC with the objective of minimizing deviations of the APC angle. This, in turn, follows the intuition that reducing this deviation decreases the risk of approaching the instability angle threshold.
    \item The post-fault APC angle and frequency of the GFM IBR are constrained by putting appropriate limits on the aforementioned MPC.
\end{enumerate}

The proposed MPC strategy is designed essentially to help the GFM IBR return to the steady-state post-fault equilibria as fast as possible. Since the complete post-fault transient behavior of the GFM IBRs is affected by both current-saturation and normal operation modes, this approach requires modeling the dynamics of the GFM IBR in both these modes, plus assess the conditions of the transitions between these two modes. Indeed, as analysed in Appendix A, the switching into the current-saturation mode happens when the APC angle exceeds a voltage-dependant threshold. The proposed MPC does not replace the APC, but corrects its trajectory by introducing optimal corrective signals. The performance of this approach is evaluated to access its limitations and capabilities in rejecting network disturbances.

The reminder of this paper is organized as follows. Section~\ref{GFM_Model} presents a model for the operation of a GFM IBR in both normal and current-saturation modes and proposes a solution for mitigating high-frequency mode oscillations. Section~\ref{TS_model} analyses the transient stability of GFM IBRs with a focus on the impact of the current saturation. This section also briefly reviews two benchmarks when analysing transient stability enhancements. Then, potential benefits of the proposed method on providing extra deceleration and modifying the APC trajectory is discussed. Section~\ref{sec:MPC_TS_Enhancement} presents the formulation of an MPC approach for achieving such transient stability enhancement. Section~\ref{sec:S_P_MPC} provides a conservative estimation of Domain of Attraction (DOA) of the post-fault transient stability of the GFM IBR controlled by the proposed method. Section~\ref{Case_Studies} analyses the performance of such a controller by means of a set of selected dynamic simulations. This is followed by a set of conclusions about the capabilities of the proposed approach in reducing the risk of transient instability and suggestions for future researches.

\section{Operational Characteristics of Grid-Forming Converter}
\label{GFM_Model}

\begin{figure*}[!htbp]
    \centering
    \includegraphics[scale=0.67]{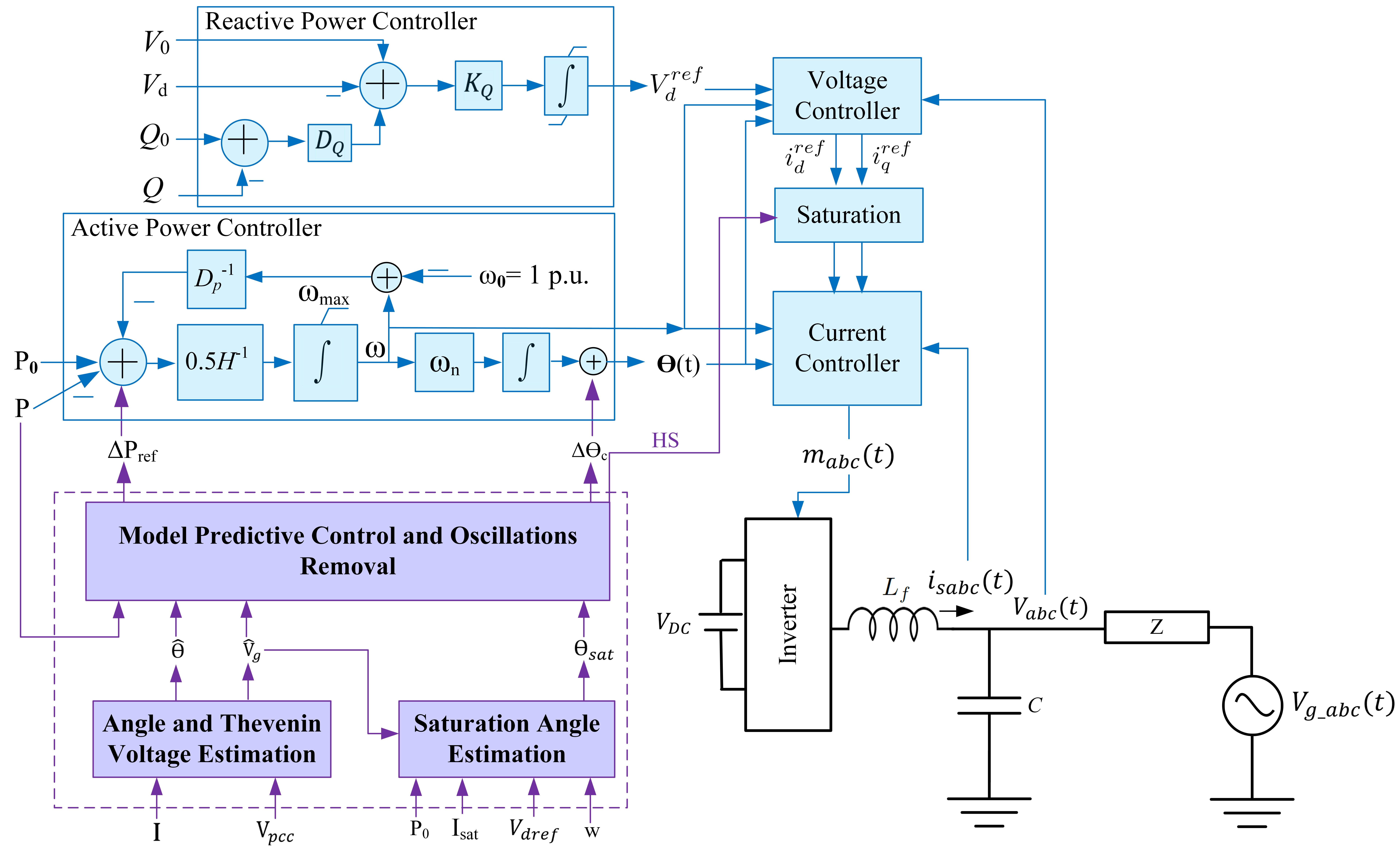}
    \caption{A block schematic viewpoint of the typical control structure of a Grid-Forming Converter with additional control structure proposed in this paper highlighted in purple.}
    \label{fig:GridForming}
\end{figure*}

This section provides an overview of (virtual synchronous machine-based) GFM IBRs \cite{Chen2022, Yazdani, TayyebiGitHub, Luo2022}. It also discusses the situations when a GFM IBR enters into the current-saturation mode or oscillates between current-saturation and normal modes.

As modeled in Fig.~\ref{fig:GridForming}, GFM IBRs are typically connected to the grid through an LC filters with inductance $L_f$ and capacitance $C$, and a transformer with impedance $Z_t$. Modeling the grid via a Thevenin equivalent, i.e., a voltage source $V_g$ behind an impedance $Z_g$ leads to an equivalent impedance of the grid plus transformer as $Z:=R+jX=Z_g+Z_t$. GFM IBRs typically consist of three layers of controllers \cite{Tayyebi2020}. The outer layer includes Reactive Power Controller (RPC) and APC. The GFM IBR synchronizes itself to the grid through APC, obtained by means of active power droop and virtual inertia \cite{Chen2022}. This also means that the arc (the argument of the cosine form the AC variable) of APC is the reference for all angles of AC variables in the dq-coordinates. To avoid confusion, recall that the arc of an AC variable is the argument of its cosine. The arc should not be confused with angle which is the relative difference between the arcs of the actual signal and the reference signal. Consider a signal $x(t) = x_m\cos \big( \omega_0 (t - t_0) + \alpha_{x0} \big)$. Angle of $x(t)$ with reference of the moment $t_0$ is $\alpha_{x0}$, which is a constant value during the steady-state. Arc of $x(t)$ is $\omega_0 (t - t_0) + \alpha_{x0}$, which always change with the rate of $\omega_0$ during the steady-state. A GFM IBR participates in reactive power sharing through the RPC,  which determines a reference $V_d^{\rm ref}$ for the inner control layer (i.e., the voltage controller). The reference $V_q^{\rm ref}$ is set to zero in the normal operation mode. In cascade, setting these references makes the voltage controller generates current references that drive the innermost controller (i.e., the current controller) in the normal operation mode.

In practice, the dynamics of the voltage and current control loops are much faster than the APC loop. This indicates that they can be neglected in the transient stability analysis \cite{Chen2022, Rokrok2022, Jin2022, Huang2019}.

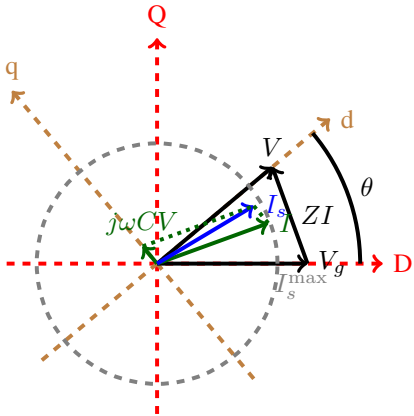
\begin{figure}[!t]
    \centering
    \scalebox{1}{\begin{tikzpicture}

    \coordinate (O) at (0,0);
    \coordinate (D) at (3,0);
    \coordinate (Q) at (0,3);
    \coordinate (d) at ($(O)!3cm!40:(D)$);
    \coordinate (q) at ($(O)!3cm!40:(Q)$);
    
    \draw[->, dashed, thick, red , line width=1.5] (-2,0) -- (D) node[anchor=west] {D};
    \draw[->, dashed, thick, red , line width=1.5] (0,-2) -- (Q) node[anchor=south] {Q};
    
    \coordinate (neg_d) at ($(O)!2cm!220:(D)$);
    \coordinate (neg_q) at ($(O)!2cm!220:(Q)$);
    \draw[->, dashed, thick, brown , line width=1.5] (neg_d) -- (d) node[anchor=west] {d};
    \draw[->, dashed, thick, brown , line width=1.5] (neg_q) -- (q) node[anchor=south] {q};

    \coordinate (V) at ($(O)!2cm!40:(D)$);
    \coordinate (Vg) at ($(O)!2cm!(D)$);
    \draw[->, thick, black , line width=1.5] (O) -- (Vg) node[at end, right] {$V_g$};
    \draw[->, thick, black , line width=1.5] (O) -- (V) node[at end, above] {$V$};
    \draw[->, thick, black , line width=1.5] (Vg) -- (V) node[midway, right] {$ZI$};

    \path let \p1=($(Vg)-(O)$), \p2=($(V)-(O)$), \n1={atan2(\y1,\x1)}, \n2 ={atan2(\y2,\x2)} in \pgfextra{\xdef\angleVg{\n1} \xdef\angleV{\n2}};

    \path let \p1=($(V)-(Vg)$), \n1={veclen(\x1,\y1)}, \n2={atan2(\y1,\x1)} in \pgfextra{\xdef\ZILength{\n1} \xdef\angleZI{\n2}};

    \coordinate (ZI_rot_start) at (O);
    \coordinate (ZI_rot_end) at ($(ZI_rot_start) + (\angleZI - 90:1.15*\ZILength)$);

    \draw[->, thick, green!40!black , line width=1.5] (ZI_rot_start) -- (ZI_rot_end) node[at end, right] {$I$};

    \coordinate (jwCV_start) at (O);
    \coordinate (jwCV_end) at ($(jwCV_start)!0.3cm!(q)$);
    \draw[->, thick, green!40!black , line width=1.5 ] (jwCV_start) -- (jwCV_end) node[at end, above] {$j\omega C V$};

    \path let \p1=($(ZI_rot_end)-(ZI_rot_start)$), \n1={\x1}, \n2={\y1} in \pgfextra{\xdef\IcompX{\n1} \xdef\IcompY{\n2}};

    \path let \p1=($(jwCV_end)-(jwCV_start)$), \n1={\x1}, \n2={\y1} in \pgfextra{\xdef\jwCVcompX{\n1} \xdef\jwCVcompY{\n2}};

    \coordinate (I_plus_jwCV_start) at (O);
    \coordinate (I_plus_jwCV_end) at ($(\IcompX+\jwCVcompX,\IcompY+\jwCVcompY)$);

    \draw[->, thick, blue , line width=1.5] (I_plus_jwCV_start) -- (I_plus_jwCV_end) node[at end,  right] {$I_s$};

    \draw[dotted , green!40!black, line width=1.5] (I_plus_jwCV_end) -- (ZI_rot_end);
    \draw[dotted, green!40!black, line width=1.5] (jwCV_end) -- (I_plus_jwCV_end);

    \def\circleRadius{1.6cm}
    \draw[dashed, gray , line width=1.5] (O) circle (\circleRadius);
    \node[anchor=north, gray , line width=1.5] at ($(O) + (-0:1.2*\circleRadius)$) {$I_s^{\mathrm{max}}$};

\draw pic["$\theta$", draw, angle radius=2.7 cm, angle eccentricity=1.1 , line width=1.5] {angle=D--O--d};

\end{tikzpicture}}
    \caption{Voltages and currents during the steady-state normal operation mode.}
    \label{fig:dqNoramlOperation}
\end{figure}

To wrap up, Fig.~\ref{fig:GridForming} shows the control diagram of a GFM IBR. From this figure, the presence of the typical virtual synchronous machine APC control with frequency bound, RPC, and internal controllers, i.e. voltage controller and current controller (in blue) are observed \cite{Chen2022, Yazdani, TayyebiGitHub, Luo2022}. The logical position of the extra controller proposed in this paper is highlighted in purple. The voltage and current phasor diagram associated with the GFM IBRs' normal operation is shown in Fig~\ref{fig:dqNoramlOperation}. Note that the Thevenin grid voltage and arc of its DQ-coordinate are virtual variables, which model the grid. The total inverter-side current ${I_s}$ is the summation of the current flowing the filter capacitor ($j\omega CV$) and the grid-side current (${I}=(V-V_g)/Z$). To prevent overheating, the inverter current should not exceed the current limit $I_s^{\mathrm{max}}$.

\subsection{The Causes and Effects of Entering into the Current-Saturation Mode}
\label{subsec:CE_Current_Saturation}
\begin{figure}[!b]
    \centering
    \scalebox{1}{\begin{tikzpicture}

    \coordinate (O) at (0,0);
    \coordinate (D) at (3,0);
    \coordinate (Q) at (0,3);
    \coordinate (d) at ($(O)!3cm!40:(D)$);
    \coordinate (q) at ($(O)!3cm!40:(Q)$);
    
    \draw[->, dashed, thick, red , line width=1.5] (-2,0) -- (D) node[anchor=west] {D};
    \draw[->, dashed, thick, red , line width=1.5] (0,-2) -- (Q) node[anchor=south] {Q};
    
    \coordinate (neg_d) at ($(O)!2cm!220:(D)$);
    \coordinate (neg_q) at ($(O)!2cm!220:(Q)$);
    \draw[->, dashed, thick, brown , line width=1.5] (neg_d) -- (d) node[anchor=west] {d};
    \draw[->, dashed, thick, brown , line width=1.5] (neg_q) -- (q) node[anchor=south] {q};

    \coordinate (V) at ($(O)!2cm!40:(D)$);
    \coordinate (Vg) at ($(O)!0.3cm!(D)$);
    \draw[->, thick, black , line width=1.5] (O) -- (Vg) node[at end, right] {$V_g$};
    \draw[->, thick, black , line width=1.5] (O) -- (V) node[at end, above] {$V$};
    \draw[->, thick, black , line width=1.5] (Vg) -- (V) node[midway, right] {$ZI$};

    \path let \p1=($(Vg)-(O)$), \p2=($(V)-(O)$), \n1={atan2(\y1,\x1)}, \n2 ={atan2(\y2,\x2)} in \pgfextra{\xdef\angleVg{\n1} \xdef\angleV{\n2}};

    \path let \p1=($(V)-(Vg)$), \n1={veclen(\x1,\y1)}, \n2={atan2(\y1,\x1)} in \pgfextra{\xdef\ZILength{\n1} \xdef\angleZI{\n2}};

    \coordinate (ZI_rot_start) at (O);
    \coordinate (ZI_rot_end) at ($(ZI_rot_start) + (\angleZI - 90:1.15*\ZILength)$);

    \draw[->, thick, green!40!black , line width=1.5] (ZI_rot_start) -- (ZI_rot_end) node[at end, above] {$I$};

    \coordinate (jwCV_start) at (O);
    \coordinate (jwCV_end) at ($(jwCV_start)!0.3cm!(q)$);
    \draw[->, thick, green!40!black , line width=1.5 ] (jwCV_start) -- (jwCV_end) node[at end, above] {$j\omega C V$};

    \path let \p1=($(ZI_rot_end)-(ZI_rot_start)$), \n1={\x1}, \n2={\y1} in \pgfextra{\xdef\IcompX{\n1} \xdef\IcompY{\n2}};

    \path let \p1=($(jwCV_end)-(jwCV_start)$), \n1={\x1}, \n2={\y1} in \pgfextra{\xdef\jwCVcompX{\n1} \xdef\jwCVcompY{\n2}};

    \coordinate (I_plus_jwCV_start) at (O);
    \coordinate (I_plus_jwCV_end) at ($(\IcompX+\jwCVcompX,\IcompY+\jwCVcompY)$);

    \draw[->, thick, blue , line width=1.5] (I_plus_jwCV_start) -- (I_plus_jwCV_end) node[at end, above ] {$I_s$};

    \draw[dotted , green!40!black, line width=1.5] (I_plus_jwCV_end) -- (ZI_rot_end);
    \draw[dotted, green!40!black, line width=1.5] (jwCV_end) -- (I_plus_jwCV_end);

    \def\circleRadius{1.6cm}
    \draw[dashed, gray , line width=1.5] (O) circle (\circleRadius);
    \node[anchor=north, gray , line width=1.5] at ($(O) + (-0:1.2*\circleRadius)$) {$I_s^{\mathrm{max}}$};


\draw pic["$\theta$", draw, angle radius=2.7 cm, angle eccentricity=1.1 , line width=1.5] {angle=D--O--d};

\end{tikzpicture}}
    \caption{An over-current caused by a \textbf{sufficiently deep voltage sag} in absence of the saturation block.}
    \label{fig:dqVoltageSagNormalOperation}
\end{figure}
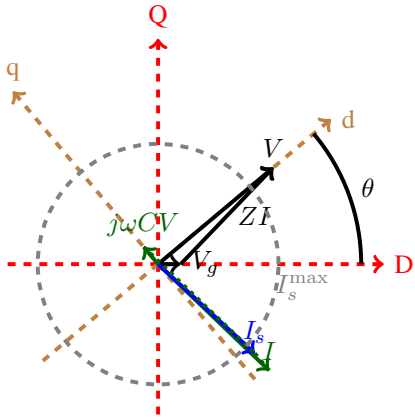

When a disturbance causes either a considerable change in the angle of the APC angle, or in the Thevenin's voltage magnitude, inverter's current may exceed the maximum limit. This is because of the fact that there is a strong link between active power with $\theta$ and reactive power with $V_g$ in an inductive transmission grid. A large APC angle results in a large active power and a considerable voltage sag leads to a large reactive power, which cause extra $I_s$ as depicted in the phasor diagrams of Fig.~\ref{fig:dqExtraAngleNormalOperation} and Fig.~\ref{fig:dqVoltageSagNormalOperation}; respectively. In that case, the native current saturation block (``Saturation'' in Fig.~\ref{fig:GridForming}) is activated. This generates a new current reference with the magnitude of $I_s^{\mathrm{max}}$ and angular displacement of $\beta$ from d-axis. As a result, the ``Voltage Controller'' in  Fig.~\ref{fig:GridForming} is deactivated.

\begin{figure}[!t]
    \centering
    \scalebox{1}{\begin{tikzpicture}

    \coordinate (O) at (0,0);
    \coordinate (D) at (3,0);
    \coordinate (Q) at (0,3);
    \coordinate (d) at ($(O)!3cm!60:(D)$);
    \coordinate (q) at ($(O)!3cm!60:(Q)$);
    
    \draw[->, dashed, thick, red, line width=1.5] (-2,0) -- (D) node[anchor=west] {D};
    \draw[->, dashed, thick, red, line width=1.5] (0,-2) -- (Q) node[anchor=south] {Q};
    
    \coordinate (neg_d) at ($(O)!2cm!240:(D)$);
    \coordinate (neg_q) at ($(O)!2cm!240:(Q)$);
    \draw[->, dashed, thick, brown, line width=1.5] (neg_d) -- (d) node[anchor=west] {d};
    \draw[->, dashed, thick, brown, line width=1.5] (neg_q) -- (q) node[anchor=south] {q};
    
    \coordinate (V) at ($(O)!2cm!60:(D)$);
    \coordinate (Vg) at ($(O)!2cm!(D)$);
    \draw[->, thick, black, line width=1.5] (O) -- (Vg) node[at end, below] {$V_g$};
    \draw[->, thick, black, line width=1.5] (O) -- (V) node[at end, above] {$V$};
    \draw[->, thick, black, line width=1.5] (Vg) -- (V) node[at end,  right] {$ZI$};

    \path let \p1=($(Vg)-(O)$), \p2=($(V)-(O)$), \n1={atan2(\y1,\x1)}, \n2 ={atan2(\y2,\x2)} in \pgfextra{\xdef\angleVg{\n1} \xdef\angleV{\n2}};

    \path let \p1=($(V)-(Vg)$), \n1={veclen(\x1,\y1)}, \n2={atan2(\y1,\x1)} in \pgfextra{\xdef\ZILength{\n1} \xdef\angleZI{\n2}};

    \coordinate (ZI_rot_start) at (O);
    \coordinate (ZI_rot_end) at ($(ZI_rot_start) + (\angleZI - 90:1.15*\ZILength)$);

    \draw[->, thick, green!40!black, line width=1.5] (ZI_rot_start) -- (ZI_rot_end) node[at end,above] {$I$};

    \coordinate (jwCV_start) at (O);
    \coordinate (jwCV_end) at ($(jwCV_start)!0.3cm!(q)$);
    \draw[->, thick, green!40!black , line width=1.5 ] (jwCV_start) -- (jwCV_end) node[at end, above] {$j\omega C V$};

    \path let \p1=($(ZI_rot_end)-(ZI_rot_start)$), \n1={\x1}, \n2={\y1} in \pgfextra{\xdef\IcompX{\n1} \xdef\IcompY{\n2}};

    \path let \p1=($(jwCV_end)-(jwCV_start)$), \n1={\x1}, \n2={\y1} in \pgfextra{\xdef\jwCVcompX{\n1} \xdef\jwCVcompY{\n2}};

     \coordinate (I_plus_jwCV_start) at (O);
    \coordinate (I_plus_jwCV_end) at ($(\IcompX+\jwCVcompX,\IcompY+\jwCVcompY)$);

    \draw[->, thick, blue , line width=1.5] (I_plus_jwCV_start) -- (I_plus_jwCV_end) node[at end, above ] {$I_s$};

    \draw[dotted , green!40!black, line width=1.5] (I_plus_jwCV_end) -- (ZI_rot_end);
    \draw[dotted, green!40!black, line width=1.5] (jwCV_end) -- (I_plus_jwCV_end);

    \def\circleRadius{1.6cm}
    \draw[dashed, gray, line width=1.5] (O) circle (\circleRadius);
    \node[anchor=north, gray] at ($(O) + (-22:1.2*\circleRadius)$) {$I_s^{\mathrm{max}}$};


\draw pic["$\theta$", draw, angle radius=2.7 cm, angle eccentricity=1.1 , line width=1.5] {angle=D--O--d};

\end{tikzpicture}}
    \caption{ An over-current caused by  \textbf{sufficiently large APC angle} $\theta$ in absence of the saturation block in} Fig.~\ref{fig:GridForming}.
    \label{fig:dqExtraAngleNormalOperation}
\end{figure}
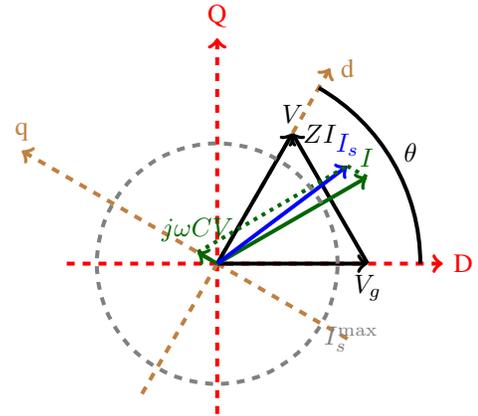

As proven in Appendix A, a GFM IBR enters into the current-saturation mode if the APC angle $\theta$ exceeds the voltage dependent threshold $\theta_{\rm sat}$ defined as

\begin{align} 
\label{eq:thetaSat}
    \cos{\theta}
    \leq
    \cfrac{1}{1-XC\omega_n\omega} 
    & (\frac{1}{2}( \cfrac{V_g}{V} + \cfrac{V}{V_g}) \\ 
    &  - \cfrac{ \left( ZI_s^{\mathrm{max}} \right)^2}{2V_g.V}    
        + \cfrac{VC^2\omega_n^2\omega^2Z^2}{2V_g} \nonumber \\
    &   - Z \cfrac{V}{V_g}C\omega_n\omega\sin{\phi} ). 
      \nonumber
\end{align} 

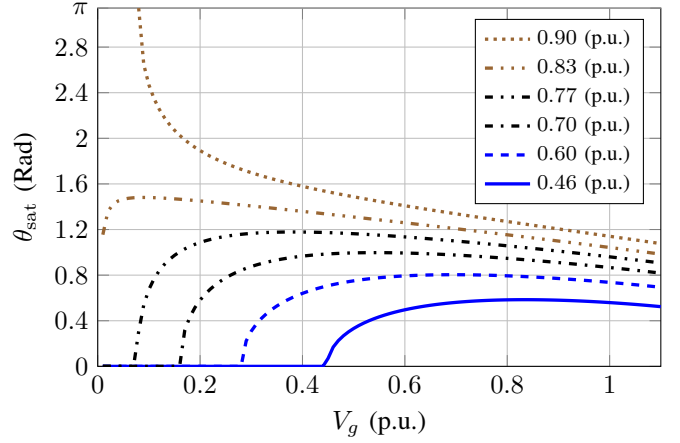
\begin{figure}[!t]
    \centering
        \begin{tikzpicture}
        \begin{axis}[
            ymax = pi,
            ymin = 0,
            xmin = 0,
            xmax = 1.1,
            xlabel={$V_g$ (p.u.)},
            ylabel={$\theta_{\rm sat}$ (Rad)},
            grid,
            ytick = {0,  0.4 ,  0.8, 1.2 , 1.6, 2.0 , 2.4 , 2.8 },
            legend pos=north east,
		reverse legend = true,
            extra y ticks={3.1415},
        extra y tick labels={$\pi$},
        extra y tick style={grid=major, major tick length=4},
        width=0.5\textwidth, height=0.35\textwidth
        ]

      \addplot[mark=none,line width=1.25, blue] table[col sep=space, x index=0, y index=1]{BodyPlot/ThetaSat.txt};
      \addlegendentry{\footnotesize $0.46$ (p.u.)}

      \addplot[dashed,line width=1.25, blue] table[col sep=space, x index=0, y index=2]{BodyPlot/ThetaSat.txt};
      \addlegendentry{\footnotesize $0.60$ (p.u.)}

      \addplot[line width=1.25, black,  dash pattern=on 3pt off 3pt on 1pt off 3pt] table[col sep=space, x index=0, y index=3]{BodyPlot/ThetaSat.txt};
      \addlegendentry{\footnotesize $0.70$ (p.u.)}  

      \addplot[dashed,line width=1.25, black ,  dash pattern= on 3pt off 3pt on 1pt off 3pt on 1pt off 3pt] table[col sep=space, x index=0, y index=4]{BodyPlot/ThetaSat.txt};
      \addlegendentry{\footnotesize $0.77$ (p.u.)} 
      
      \addplot[line width=1.25, brown!80!black, dash pattern=on 3pt off 3pt on 1pt off 3pt on 1pt off 3pt on 1pt off 3pt] table[col sep=space, x index=0, y index=5]{BodyPlot/ThetaSat.txt};
      \addlegendentry{\footnotesize $0.83 $ (p.u.)} 

      \addplot[dashed,line width=1.25, brown!80!black, dotted] table[col sep=space, x index=0, y index=1]{BodyPlot/ThetaSatWeak.txt};
      \addlegendentry{\footnotesize $0.90$  (p.u.)} 
      
        \end{axis}
    \end{tikzpicture}
    \caption{Variations of $\theta_{\rm sat}$ with the grid Thevenin's voltage $V_g$ for different values of equivalent impedance ($Z$) specified in the legend.}
    \label{fig:ThetaSat}
\end{figure}

Note that, neglecting the effect of the filter's capacitance (it is small and has negligible impact on the actual voltages and currents), such threshold depends on the grid strength and grid's Thevenin's voltage (as summarized in Fig.~\ref{fig:ThetaSat}). More verbosely, for a strong grid, the saturation APC angle threshold is relatively smaller specially for a deep voltage sag. However, as the grid becomes weaker, the saturation threshold $\theta_{\rm sat}$ becomes larger. This is because of the fact that a smaller current ($I$) in Fig.~\ref{fig:dqNoramlOperation} yields the same $ZI$ if the grid is weaker (larger $Z$). Therefore, a larger vector difference between $V$ and $V_g$ can be accommodated by the allowed current in the normal operation mode when connected to a weak grid. Fig.~\ref{fig:ThetaSat} depicts that if the grid is exceedingly weak, a severe voltage sag does not cause the GFM IBR fall into the current-saturation mode.

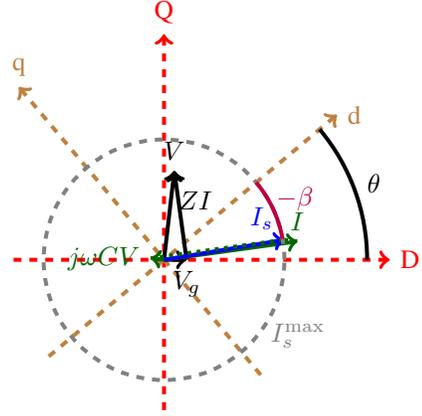
\begin{figure}[!t]
    \centering
    \scalebox{1}{\begin{tikzpicture}

    \coordinate (O) at (0,0);
    \coordinate (D) at (3,0);
    \coordinate (Q) at (0,3);
    \coordinate (d) at ($(O)!3cm!40:(D)$);
    \coordinate (q) at ($(O)!3cm!40:(Q)$);
    
\draw[->, dashed, thick, red, line width=1.5] (-2,0) -- (D) node[anchor=west] {D};
    \draw[->, dashed, thick, red, line width=1.5] (0,-2) -- (Q) node[anchor=south] {Q};
    
    \coordinate (neg_d) at ($(O)!2cm!220:(D)$);
    \coordinate (neg_q) at ($(O)!2cm!220:(Q)$);
    \draw[->, dashed, thick, brown , line width=1.5] (neg_d) -- (d) node[anchor=west] {d};
    \draw[->, dashed, thick, brown , line width=1.5] (neg_q) -- (q) node[anchor=south] {q};
    
    \coordinate (Vg) at ($(O)!0.3cm!(D)$);

    \coordinate (V) at (0.1366 , 1.1818);

    \draw[->, thick, black , line width=1.5] (O) -- (Vg) node[at end, below] {$V_g$};    
    \draw[->, thick, black , line width=1.5] (O) -- (V) node[at end, above ] {$V$};
    \draw[->, thick, black , line width=1.5] (Vg) -- (V) node[pos=0.5, xshift=2mm, yshift=2mm] {$ZI$};

    \path let \p1=($(Vg)-(O)$), \p2=($(V)-(O)$), \n1={atan2(\y1,\x1)}, \n2 ={atan2(\y2,\x2)} in \pgfextra{\xdef\angleVg{\n1} \xdef\angleV{\n2}};

    \path let \p1=($(V)-(Vg)$), \n1={veclen(\x1,\y1)}, \n2={atan2(\y1,\x1)} in \pgfextra{\xdef\ZILength{\n1} \xdef\angleZI{\n2}};

    \coordinate (ZI_rot_start) at (O);
    \coordinate (ZI_rot_end) at ($(ZI_rot_start) + (\angleZI - 90:1.5*\ZILength)$);

    \draw[->, thick, green!40!black  , line width=1.5] (ZI_rot_start) -- (ZI_rot_end) node[at end,above] {$I$};

    \coordinate (jwCV_start) at (O);

        
    \coordinate (jwCV_end) at (-0.1773 , 0.0205);

    \draw[->, thick, green!40!black , line width=1.5] (jwCV_start) -- (jwCV_end) node[at end, left] {$j\omega C V$};

    \path let \p1=($(ZI_rot_end)-(ZI_rot_start)$), \n1={\x1}, \n2={\y1} in \pgfextra{\xdef\IcompX{\n1} \xdef\IcompY{\n2}};

    \path let \p1=($(jwCV_end)-(jwCV_start)$), \n1={\x1}, \n2={\y1} in \pgfextra{\xdef\jwCVcompX{\n1} \xdef\jwCVcompY{\n2}};

    \coordinate (I_plus_jwCV_start) at (O);
    \coordinate (I_plus_jwCV_end) at ($(\IcompX+\jwCVcompX,\IcompY+\jwCVcompY)$);

    \draw[->, thick, blue , line width=1.5] (I_plus_jwCV_start) -- (I_plus_jwCV_end) node[at end, above left] {$I_s$};

    \draw[dotted , green!40!black, line width=1.5] (I_plus_jwCV_end) -- (ZI_rot_end);
    \draw[dotted, green!40!black, line width=1.5] (jwCV_end) -- (I_plus_jwCV_end);
    
    \def\circleRadius{1.6cm}
    \draw[dashed, gray , line width=1.5] (O) circle (\circleRadius);
    \node[anchor=north, gray] at ($(O) + (-22:1.2*\circleRadius)$) {$I_s^{\mathrm{max}}$};

    \def\betaa{-30}
    
    \coordinate (beta) at ($(\betaa+40:1.6cm)$);

    \draw pic["$-\beta$", draw, angle radius=1.6 cm, angle eccentricity=1.2, purple , line width=1.5] {angle=beta--O--d};


\draw pic["$\theta$", draw, angle radius=2.7 cm, angle eccentricity=1.1 , line width=1.5] {angle=D--O--d};

\end{tikzpicture}}
    \caption{Voltages and currents in the current-saturation mode induced by \textbf{a deep voltage sag}.}
    \label{fig:dqVoltageSagSaturation}
\end{figure}

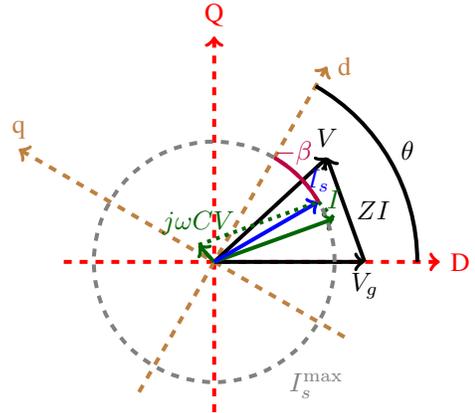
\begin{figure}[!t]
    \centering
    \scalebox{1}{\begin{tikzpicture}

    \coordinate (O) at (0,0);
    \coordinate (D) at (3,0);
    \coordinate (Q) at (0,3);
    \coordinate (d) at ($(O)!3cm!60:(D)$);
    \coordinate (q) at ($(O)!3cm!60:(Q)$);
    
    \draw[->, dashed, thick, red, line width=1.5] (-2,0) -- (D) node[anchor=west] {D};
    \draw[->, dashed, thick, red, line width=1.5] (0,-2) -- (Q) node[anchor=south] {Q};
    
    \coordinate (neg_d) at ($(O)!2cm!240:(D)$);
    \coordinate (neg_q) at ($(O)!2cm!240:(Q)$);
    \draw[->, dashed, thick, brown, line width=1.5] (neg_d) -- (d) node[anchor=west] {d};
    \draw[->, dashed, thick, brown, line width=1.5] (neg_q) -- (q) node[anchor=south] {q};
    
    \coordinate (V) at (1.5, 1.3856);

    \coordinate (Vg) at ($(O)!2cm!(D)$);
    \draw[->, thick, black, line width=1.5] (O) -- (Vg) node[at end, below] {$V_g$};
    \draw[->, thick, black, line width=1.5] (O) -- (V) node[at end, above] {$V$};
    \draw[->, thick, black, line width=1.5] (Vg) -- (V) node[pos=0.5, right] {$ZI$};

    \path let \p1=($(Vg)-(O)$), \p2=($(V)-(O)$), \n1={atan2(\y1,\x1)}, \n2 ={atan2(\y2,\x2)} in \pgfextra{\xdef\angleVg{\n1} \xdef\angleV{\n2}};

    \path let \p1=($(V)-(Vg)$), \n1={veclen(\x1,\y1)}, \n2={atan2(\y1,\x1)} in \pgfextra{\xdef\ZILength{\n1} \xdef\angleZI{\n2}};

    \coordinate (ZI_rot_start) at (O);
    \coordinate (ZI_rot_end) at ($(ZI_rot_start) + (\angleZI - 90:1.15*\ZILength)$);

    \draw[->, thick, green!40!black, line width=1.5] (ZI_rot_start) -- (ZI_rot_end) node[at end,above] {$I$};

    \coordinate (jwCV_start) at (O);
    \coordinate (jwCV_end) at (-0.2078 , 0.2250);
    \draw[->, thick, green!40!black , line width=1.5] (jwCV_start) -- (jwCV_end) node[at end, above] {$j\omega C V$};

    \path let \p1=($(ZI_rot_end)-(ZI_rot_start)$), \n1={\x1}, \n2={\y1} in \pgfextra{\xdef\IcompX{\n1} \xdef\IcompY{\n2}};

    \path let \p1=($(jwCV_end)-(jwCV_start)$), \n1={\x1}, \n2={\y1} in \pgfextra{\xdef\jwCVcompX{\n1} \xdef\jwCVcompY{\n2}};

     \coordinate (I_plus_jwCV_start) at (O);
    \coordinate (I_plus_jwCV_end) at ($(\IcompX+\jwCVcompX,\IcompY+\jwCVcompY)$);

    \draw[->, thick, blue , line width=1.5] (I_plus_jwCV_start) -- (I_plus_jwCV_end) node[at end, above ] {$I_s$};

    \draw[dotted , green!40!black, line width=1.5] (I_plus_jwCV_end) -- (ZI_rot_end);
    \draw[dotted, green!40!black, line width=1.5] (jwCV_end) -- (I_plus_jwCV_end);

    \def\circleRadius{1.6cm}
    \draw[dashed, gray , line width=1.5] (O) circle (\circleRadius);
    \node[anchor=north, gray] at ($(O) + (-45:1.2*\circleRadius)$) {$I_s^{\mathrm{max}}$};

    \def\betaa{-30}
    
    \coordinate (beta) at ($(\betaa+60:1.95cm)$);

    \draw pic["$-\beta$", draw, angle radius=1.6 cm, angle eccentricity=1.28, purple, line width=1.5, left] {angle=beta--O--d};


\draw pic["$\theta$", draw, angle radius=2.7 cm, angle eccentricity=1.1, line width=1.5] {angle=D--O--d};

\end{tikzpicture}}
    \caption{Voltages and currents in the current-saturation mode induced by \textbf{large APC angle $\theta$}.}
    \label{fig:dqExtraAngleSaturationOperationMode}
\end{figure}

As mentioned earlier in this section, current saturation causes deactivation of the current reference generated by the voltage controller. Hence, the GFM IBR's voltage $V$ in this case does not follow its reference ($V_d^{\rm ref}$, $V_q^{\rm ref}$), and it is determined by $V_g$, $I_s^{\mathrm{max}}$, $\beta$, $\theta$, and $Z$ as depicted in Fig.~\ref{fig:dqVoltageSagSaturation} and Fig.~\ref{fig:dqExtraAngleSaturationOperationMode}.  Neglecting again the impact of the filter capacitors, one may perform a vector analysis as in Appendix B, and obtain that the voltage under the current-saturation mode evolves as
\begin{align}
    V_d^{\rm sat}
    & =
    V_g\cos{\theta}
    + ZI_s^{\mathrm{max}}\cos{(\beta+\phi)}
    \label{eq:Vdsat}, \\
    V_q^{\rm sat}
    & =
    -V_g\sin{\theta}
    + ZI_s^{\mathrm{max}}\sin{(\beta+\phi)} .
    \label{eq:Vqsat}
\end{align}

\subsection{Mitigation of Mode Oscillations During Post-fault Recovery}
\label{subsec:Oscillations}

While in the current-saturation mode, the voltage controller is deactivated, but still generates a current reference. If this reference during the current-saturation operation mode is larger than the maximum allowed current, then the GFM IBR does not switch back to the normal mode \cite{Fan2022}. The magnitude of this reference in its turn depends on the GFM IBR's APC angle, the parameters of voltage controller, the anti wind-up strategy, as well as the post-fault grid conditions. In some certain situations, this reference becomes less than the allowed value, and the GFM IBR returns to a normal mode, but the conditions after entering the normal mode are such that it immediately re-enters in saturation. In these situations, the GFM IBR switches back-and-forth between current-saturation and normal modes, inducing oscillations in the system \cite{Fan2022}.

The presence of mode oscillations introduces a high-frequency disturbance for a couple of hundreds of milliseconds. To mitigate this phenomenon, this paper proposes to hold the GFM IBR in the current-saturation mode if the cosine of the APC angle $\theta$ is less than 95\% of the threshold introduced in ~\eqref{eq:thetaSat} , regardless of the magnitude of the current reference produced by the voltage controller. In case of errors in estimating variables and parameters, the criteria in ~\eqref{eq:thetaSat} might have error. This error causes holding in saturation unnecessarily which compromise the transient stability. To prevent enforcing saturation more than the required one, only up to 95\% of the criteria saturation is enforced to remain in saturation. Therefore, the minor oscillations might still exist. Nevertheless, a huge amount of oscillations is mitigated by holding saturation.

\section{Transient Stability of Grid-Forming Converter}
\label{TS_model}

The concept of transient stability of a GFM IBR relates to its ability to retain synchronism after a large disturbance \cite{Luo2022}. Synchronism is mainly associated with the APC angle and frequency.
Based on the control design shown in Fig.~\ref{fig:GridForming}, the APC loop follows the swing equations~\cite{Kundur}, i.e.,
\begin{align}
    P_0-P
    &=
    2H \cfrac{d\omega}{dt}
    +
    \cfrac{1}{D_p} \left( \omega - \omega_0 \right) , 
    \label{swing}
    \\
    \cfrac{d{\theta}}{dt}
    &=
    \omega_n \left( \omega - \omega_0 \right) \; .
    \label{thetaOmega}
\end{align}
Neglecting the effect of the damping term $D_p$  enables performing a transient angle stability analysis of GFM IBRs by means of the traditional \emph{equal area criterion} \cite{Kundur}. However, because of the current saturation, the relation between output power and APC angle changes. Indeed, the power-angle relations during normal and current-saturation modes can be expressed, respectively, as \cite{Rokrok2022} and Appendix B
\begin{align}
P_{\rm unsat} &= \cfrac{V_g V}{X}\sin{\theta}\\
P_{\rm sat} &= \cfrac{I_s^{\mathrm{max}}}{1-\omega_0\omega_nCX}V_g\cos{(\theta + \beta)} \; .
\end{align}
These equations are decent approximations only if the resistive component of $Z$ is insignificant. According to these equations, the saturated power output is considerably less dependent on $X$ if filter capacitor is negligible.

\begin{figure}[!t]
    \centering
    \includegraphics[scale=0.3]{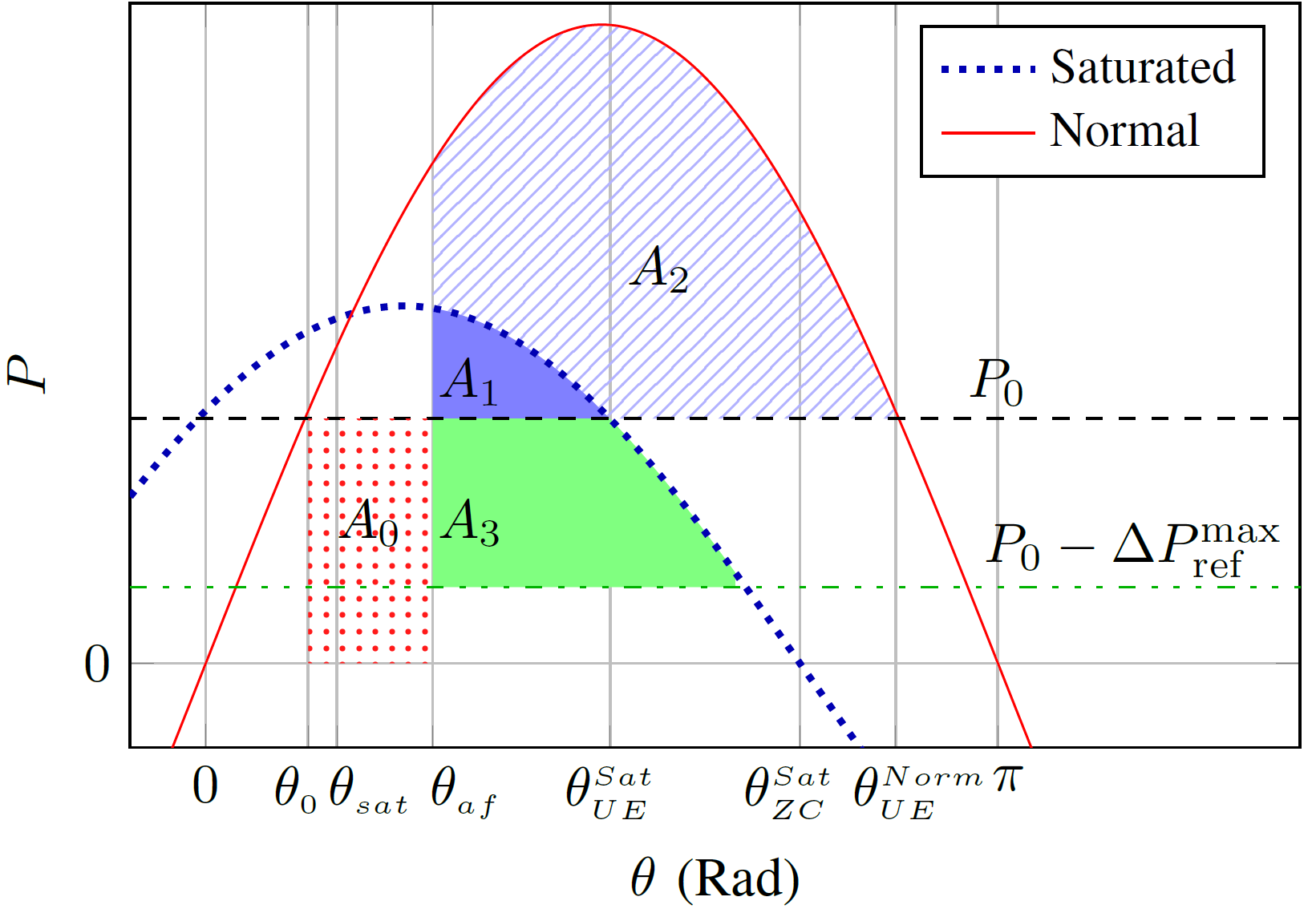}
    \caption{Power-angle curve.}
    \label{fig:SatandUnsatPower}
\end{figure}

To graphically compare the effect of normal and saturated output powers on transient stability, Fig.~\ref{fig:SatandUnsatPower} shows how the deceleration area for the saturated power output (area $A_1$) is considerably smaller than the unsaturated deceleration one (area $A_1$ plus area $A_2$). Here $\theta_0$ is the initial APC angle, $\theta_{\rm af}$ is the APC angle immediately after the fault is cleared, $\theta_{\rm sat}$ is the saturation threshold. Since active power is not the only component of the apparent power which makes the output current, $\theta_{\rm sat}$ is not the intersection of active saturated power and normal power curves. The effect of reactive power leads to a minor deviation of $\theta_{\rm sat}$ from this intersection point. $\theta_{\rm UE}^{\rm sat}$ and $\theta_{\rm ZC}^{\rm sat}$ are the APC angles corresponding to the saturated unstable equilibrium angle and zero power output angle (zero crossing angle), respectively. These two angles are in their turn smaller than those of the unsaturated case (i.e., $\theta_{\rm UE}^{\rm Norm}$ and $\theta_{\rm ZC}^{\rm unsat}=\pi$). To summarize, entering the current-saturation mode reduces the deceleration area, the unstable equilibrium angle $\theta_{\rm UE}$ and the zero power angle $\theta_{\rm ZC}$.

Once the APC angle exceeds $\theta_{\rm UE}^{sat}$, the GFM IBR will be unable to return to the stable equilibrium without any corrective control. Adopting an effective corrective control, which will be explained in more details in Subsection~\ref{Solution}, means steering the APC angle and frequency so that even if the APC angle exceeds $\theta_{\rm UE}^{\rm sat}$ the system can be brought back to the Stable Equilibrium Point (SEP). However, after passing $\theta_{\rm ZC}^{\rm sat}$ from which the GFM IBR may absorb power reversely, the system is being operated in an unsafe condition and is considered unsafely unstable. Operating GFM IBR in APC angles less than $\theta_{\rm ZC}^{\rm sat}$ is considered \textit{safe} because there is a chance for the inverter to return to the normal operation mode by employing an appropriate and safe corrective control before it absorbs power from the grid. In Section~\ref{sec:MPC_TS_Enhancement}, it will be discussed how safety is ensured as a constraint of the proposed MPC in addition to stability.

\begin{figure}[!t]
    \centering
        \begin{tikzpicture}
        \begin{axis}[
            ymax = 0.05,
            ymin = -0.03,
            xmax = pi,
            xmin = 0,
            xlabel={$\theta$},
            ylabel={$\Delta\omega$ (p.u.)},
            grid,
            legend pos=south east,
            extra x ticks={0.5593, 1.545, 2.7294},
            extra x tick labels={$\theta_{\scriptscriptstyle \rm sat}$, $\theta_{\scriptscriptstyle \rm UE}^{\scriptscriptstyle \rm Sat}$, $\theta_{\scriptscriptstyle \rm UE}^{\scriptscriptstyle \rm Unsat}$},
            grid,
            xticklabels={},
            xtick={0}, 
            width=0.5\textwidth, height=0.30\textwidth
        ]

       \addplot[mark=none,line width=1.25, blue] table[col sep=space,  x expr=\thisrowno{0}, y index=1]{BodyPlot/Trajectories/BoundarySat.txt};
       \addlegendentry{\footnotesize  Sat.}

       \addplot[mark=none,line width=1.25, brown] table[col sep=space,  x expr=\thisrowno{0}, y index=1]{BodyPlot/Trajectories/BoundaryUnsat.txt};
       \addlegendentry{\footnotesize  Unsat.}

      \addplot[mark=none,line width=0.75, black , dashed , postaction={ decorate, decoration={ markings, mark=at position 0.25 with {\arrow{stealth}}}}] table[col sep=space,  x expr=\thisrowno{8}, y index=9]{BodyPlot/Trajectories/SatTrajectories.txt};

     \addplot[mark=none,line width=0.75, black , dashed , postaction={ decorate, decoration={ markings, mark=at position 0.25 with {\arrow{stealth}}}}] table[col sep=space,  x expr=\thisrowno{18}, y index=19]{BodyPlot/Trajectories/SatTrajectories.txt};

     \addplot[mark=none,line width=0.75, black , dashed , postaction={ decorate, decoration={ markings, mark=at position 0.25 with {\arrow{stealth}}}}] table[col sep=space,  x expr=\thisrowno{12}, y index=13]{BodyPlot/Trajectories/SatTrajectories.txt};

    \addplot[mark=none,line width=0.75, black , dashed , postaction={ decorate, decoration={ markings, mark=at position 0.25 with {\arrow{stealth}}}}] table[col sep=space,  x expr=\thisrowno{16}, y index=17]{BodyPlot/Trajectories/SatTrajectories.txt};

    \addplot[mark=none,line width=0.75, black , dashed , postaction={ decorate, decoration={ markings, mark=at position 0.25 with {\arrow{stealth}}}}] table[col sep=space,  x expr=\thisrowno{28}, y index=29]{BodyPlot/Trajectories/SatTrajectories.txt};  

    \addplot[mark=none,line width=0.75, black , dashed , postaction={ decorate, decoration={ markings, mark=at position 0.25 with {\arrow{stealth}}}}] table[col sep=space,  x expr=\thisrowno{40}, y index=41]{BodyPlot/Trajectories/SatTrajectories.txt}; 

    \addplot[mark=none,line width=0.75, black , dashed , postaction={ decorate, decoration={ markings, mark=at position 0.25 with {\arrow{stealth}}}}] table[col sep=space,  x expr=\thisrowno{46}, y index=47]{BodyPlot/Trajectories/SatTrajectories.txt};   

    \addplot[mark=none,line width=0.75, black , dashed , postaction={ decorate, decoration={ markings, mark=at position 0.21 with {\arrow{stealth}}}}] table[col sep=space,  x expr=\thisrowno{52}, y index=53]{BodyPlot/Trajectories/SatTrajectories.txt};

      \addplot[mark=none,line width=0.75, green!50!black , dashed , postaction={ decorate, decoration={ markings, mark=at position 0.25 with {\arrow{stealth}}}}] table[col sep=space,  x expr=\thisrowno{0}, y index=1]{BodyPlot/Trajectories/UnsatTrajectories.txt};

      \addplot[mark=none,line width=0.75, green!50!black , dashed , postaction={ decorate, decoration={ markings, mark=at position 0.25 with {\arrow{stealth}}}}] table[col sep=space,  x expr=\thisrowno{4}, y index=5]{BodyPlot/Trajectories/UnsatTrajectories.txt};
      
      \addplot[mark=none,line width=0.75, green!50!black , dashed , postaction={ decorate, decoration={ markings, mark=at position 0.25 with {\arrow{stealth}}}}] table[col sep=space,  x expr=\thisrowno{6}, y index=7]{BodyPlot/Trajectories/UnsatTrajectories.txt};

      \addplot[mark=none,line width=0.75, green!50!black , dashed , postaction={ decorate, decoration={ markings, mark=at position 0.25 with {\arrow{stealth}}}}] table[col sep=space,  x expr=\thisrowno{8}, y index=9]{BodyPlot/Trajectories/UnsatTrajectories.txt};      

      \addplot[mark=none,line width=0.75, green!50!black , dashed , postaction={ decorate, decoration={ markings, mark=at position 0.25 with {\arrow{stealth}}}}] table[col sep=space,  x expr=\thisrowno{10}, y index=11]{BodyPlot/Trajectories/UnsatTrajectories.txt};

     \addplot[mark=none,line width=0.75, green!50!black , dashed , postaction={ decorate, decoration={ markings, mark=at position 0.25 with {\arrow{stealth}}}}] table[col sep=space,  x expr=\thisrowno{14}, y index=15]{BodyPlot/Trajectories/UnsatTrajectories.txt};

     \addplot[mark=none,line width=0.75, green!50!black , dashed , postaction={ decorate, decoration={ markings, mark=at position 0.25 with {\arrow{stealth}}}}] table[col sep=space,  x expr=\thisrowno{16}, y index=17]{BodyPlot/Trajectories/UnsatTrajectories.txt};

     \addplot[mark=none,line width=0.75, green!50!black , dashed , postaction={ decorate, decoration={ markings, mark=at position 0.25 with {\arrow{stealth}}}}] table[col sep=space,  x expr=\thisrowno{18}, y index=19]{BodyPlot/Trajectories/UnsatTrajectories.txt};

     \addplot[mark=none,line width=0.75, green!50!black , dashed , postaction={ decorate, decoration={ markings, mark=at position 0.25 with {\arrow{stealth}}}}] table[col sep=space,  x expr=\thisrowno{14}, y index=15]{BodyPlot/Trajectories/UnsatTrajectories.txt};  

    \addplot[mark=none,line width=0.75, green!50!black , dashed , postaction={ decorate, decoration={ markings, mark=at position 0.25 with {\arrow{stealth}}}}] table[col sep=space,  x expr=\thisrowno{16}, y index=17]{BodyPlot/Trajectories/UnsatTrajectories.txt};  

    \addplot[mark=none,line width=0.75, green!50!black , dashed , postaction={ decorate, decoration={ markings, mark=at position 0.25 with {\arrow{stealth}}}}] table[col sep=space,  x expr=\thisrowno{20}, y index=21]{BodyPlot/Trajectories/UnsatTrajectories.txt};  

    \addplot[mark=none,line width=0.75, green!50!black , dashed , postaction={ decorate, decoration={ markings, mark=at position 0.25 with {\arrow{stealth}}}}] table[col sep=space,  x expr=\thisrowno{28}, y index=29]{BodyPlot/Trajectories/UnsatTrajectories.txt};  

    \addplot[mark=none,line width=0.75, green!50!black , dashed , postaction={ decorate, decoration={ markings, mark=at position 0.25 with {\arrow{stealth}}}}] table[col sep=space,  x expr=\thisrowno{34}, y index=35]{BodyPlot/Trajectories/UnsatTrajectories.txt}; 

    \addplot[mark=none,line width=0.75, green!50!black , dashed , postaction={ decorate, decoration={ markings, mark=at position 0.25 with {\arrow{stealth}}}}] table[col sep=space,  x expr=\thisrowno{40}, y index=41]{BodyPlot/Trajectories/UnsatTrajectories.txt};

        \end{axis}
    \end{tikzpicture}
    \caption{Boundaries of DOA and post-fault trajectories for a GFM IBR equipped with CRS (blue line and black lines) and not equipped with CRS (brown line and green lines).}
    \label{fig:DOASatUnsat}
\end{figure}
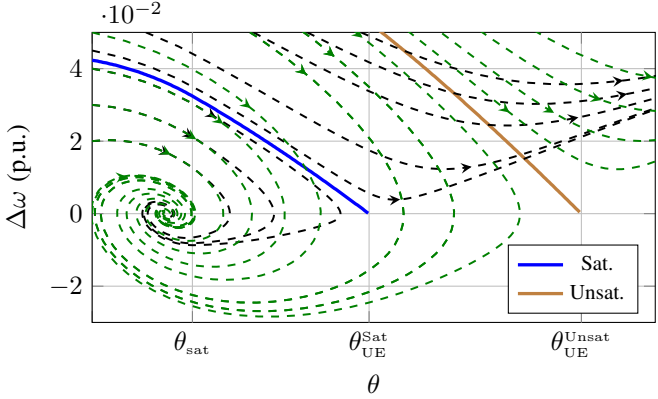

Fig.~\ref{fig:DOASatUnsat} shows boundaries of post-fault stability of GFM IBR for two cases \emph{a)} in which it enters current-saturation mode in case of an over-current and \emph{b)} it has huge over-current capacity and is not saturated. Trajectories for saturated and unsaturated cases are numerically calculated and shown with dashed black and green lines respectively. Before reaching saturation angle, trajectories of both cases are identical. This figure also confirms that the saturation deteriorates transient stability margin.

\subsection{Benchmarks in Transient Stability Enhancement}\label{benchmark}

Two existing control strategies are used for comparison: (1) \emph{Bounding the Frequency} \cite{Luo2022} and (2) \emph{Compensating for Saturation} \cite{Kkuni}.

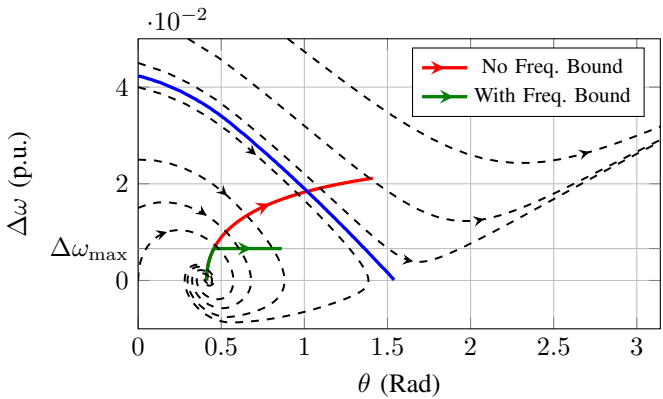
\begin{figure}[!t]
    \centering
        \begin{tikzpicture}
        \begin{axis} [
            ymax = 0.05,
            ymin = -0.01,
            xmax = pi,
            xmin = 0,
            xlabel={$\theta$ (Rad)},
            ylabel={$\Delta\omega$ (p.u.)},
            grid,
            extra y ticks={0.0066},
            extra y tick labels={$\Delta\omega_{\rm max}$},
            legend pos=north east,
            width=0.47\textwidth, height=0.30\textwidth
        ]

       \addplot[mark=none,line width=1.25, red , postaction={ decorate, decoration={ markings, mark=at position 0.5 with {\arrow{stealth}}}}] table[col sep=space,  x expr=\thisrowno{0}, y index=1]{BodyPlot/Trajectories/NoFreqBoundFaultOn.txt};
       \addlegendentry{\footnotesize No Freq. Bound} 

       \addplot[mark=none,line width=1.25, green!50!black, postaction={ decorate, decoration={ markings, mark=at position 0.7 with {\arrow{stealth}}}}] table[col sep=space,  x expr=\thisrowno{0}, y index=1]{BodyPlot/Trajectories/FreqBoundFaultOn.txt};
       \addlegendentry{\footnotesize With Freq. Bound}

      \addplot[mark=none,line width=0.75, black , dashed , postaction={ decorate, decoration={ markings, mark=at position 0.25 with {\arrow{stealth}}}}] table[col sep=space,  x expr=\thisrowno{0}, y index=1]{BodyPlot/Trajectories/SatTrajectories.txt};

      \addplot[mark=none,line width=0.75, black , dashed , postaction={ decorate, decoration={ markings, mark=at position 0.25 with {\arrow{stealth}}}}] table[col sep=space,  x expr=\thisrowno{6}, y index=7]{BodyPlot/Trajectories/SatTrajectories.txt};

      \addplot[mark=none,line width=0.75, black , dashed , postaction={ decorate, decoration={ markings, mark=at position 0.25 with {\arrow{stealth}}}}] table[col sep=space,  x expr=\thisrowno{10}, y index=11]{BodyPlot/Trajectories/SatTrajectories.txt};

     \addplot[mark=none,line width=0.75, black , dashed , postaction={ decorate, decoration={ markings, mark=at position 0.25 with {\arrow{stealth}}}}] table[col sep=space,  x expr=\thisrowno{16}, y index=17]{BodyPlot/Trajectories/SatTrajectories.txt};

     \addplot[mark=none,line width=0.75, black , dashed , postaction={ decorate, decoration={ markings, mark=at position 0.25 with {\arrow{stealth}}}}] table[col sep=space,  x expr=\thisrowno{18}, y index=19]{BodyPlot/Trajectories/SatTrajectories.txt};

    \addplot[mark=none,line width=0.75, black , dashed , postaction={ decorate, decoration={ markings, mark=at position 0.25 with {\arrow{stealth}}}}] table[col sep=space,  x expr=\thisrowno{28}, y index=29]{BodyPlot/Trajectories/SatTrajectories.txt};

    \addplot[mark=none,line width=0.75, black , dashed , postaction={ decorate, decoration={ markings, mark=at position 0.25 with {\arrow{stealth}}}}] table[col sep=space,  x expr=\thisrowno{40}, y index=41]{BodyPlot/Trajectories/SatTrajectories.txt};

       \addplot[mark=none,line width=1.25, blue] table[col sep=space,  x expr=\thisrowno{0}, y index=1]{BodyPlot/Trajectories/BoundarySat.txt};

    \end{axis}
    \end{tikzpicture}
    \caption{Fault-on trajectories of a GFM IBR (\emph{a}) equipped and (\emph{b}) not equipped with frequency bounding in case $X = 0.46$ p.u. The blue line is the boundary of stable region.}
    \label{fig:FreqBoundEffect}
\end{figure}

\emph{Bounding the Frequency} essentially corresponds to limit the acceleration area in Fig.~\ref{fig:SatandUnsatPower} by limiting the frequency produced by APC. This limitation contributes to transient stability enhancement in two ways. First of all, the post-fault frequency is smaller, meaning that less deceleration is needed to bring the frequency back to the normal operating frequency. Secondly, the post-fault APC angle ($\theta_{af}$ in Fig.~\ref{fig:SatandUnsatPower}) becomes considerably smaller than when operating in the unbounded cases as also shown in Fig.~\ref{fig:FreqBoundEffect}. In this figure, trajectories are shown with dashed lines. By bounding the frequency, the APC angle becomes less than the stability boundary unless the fault duration is too long. Note that the boundary (maximum) frequency should be sufficiently large so that the GFM IBR can participate in active power sharing during the normal operation. The smaller this frequency limit, the longer duration of a fault is stable; however, the GFM will be limited in participating grid frequency control.

\emph{Compensating for Saturation} aims at mitigating the adverse effect of saturation on transient stability by means of subtracting a virtual power from the reference power in the APC loop. As suggested in \cite{Kkuni}, to compute such a virtual power one may use the unsaturated current reference, a method though only applicable for some specific voltage controllers (e.g., a virtual impedance voltage controller) considered in \cite{Kkuni}. For PI voltage controller (which is used in this paper), current saturation makes the voltage controller deactivated as explained in Subsection~\ref{subsec:Oscillations}. Therefore, the current reference produced by the voltage controller is not a meaningful signal in the control scheme of Fig.~\ref{fig:GridForming} during the current-saturation mode operation. In this paper, in order to adopt this approach as a benchmark (not as the main proposed methodology) for assessing our proposed method, the difference between unsaturated and saturated power is first computed based on the relation between the active power and estimated APC angle, and then subtracted from the reference power. This approach provides extra post-fault deceleration compared to the uncompensated control. 

This extra deceleration could be close to the required one. However, if it is smaller than the required amount, it does not help the GFM IBR become stable, while if it is larger, this might lead to additional frequency and angle oscillations (vanishing in time as the system settles to its steady-state equilibrium).

\subsection{Proposed Solution}\label{Solution}

As shown in Fig.~\ref{fig:DOASatUnsat} the current saturation considerably shrinks the DOA. The post-fault APC trajectories can be barely modified by changing the control parameters. However, adjusting the parameters or restructuring the APC controller so to enhance transient stability might compromise several stability properties, e.g., the small-signal stability in some conditions. Therefore, we are interested in investigating an approach that does not rely in modifying the APC structure and/or its parameters. Instead, we are interested in checking the capabilities of triggering, when entering in a post-fault current-saturation mode, an MPC scheme whose aim is to introduce a corrective phase jump, and to change the reference power so to redirect the APC trajectory ($\theta$ and $\Delta \omega$).

Indeed the proposed MPC-based scheme may provide the transient stability of GFM IBR with the following opportunities:

\begin{itemize}

\item An enhanced controllability over the APC's state variables ($\theta$, $\Delta \omega$ ) and thus trajectory. Indeed, dynamically changing the reference power $\Delta P_{\rm ref}$ induces changes in $\Delta \omega$, and as a result a change in $\theta$. Introducing a corrective phase jump $\Delta \theta_c$ also directly changes $\theta$. Hence, an optimization solver may have the capability of optimizing the system's trajectory by suitably evaluating these two decision variables (control actuators). 

\item Moreover an MPC not only increases the chance of reaching SEP, but also can make use of opportune constraints on transient operation of GFM IBR, and thus act according to safety criteria and operator's desires.

\end{itemize}

The extra deceleration area $A_3$ in Fig.~\ref{fig:SatandUnsatPower} is provided if the reference power changes from $P_0$ to $P_0-\Delta P_{\rm ref}^{\rm max}$. Besides, the APC angle can directly turn back to the stable operating area by using corrective phase jumps. In other words, these two signals provides the opportunity to correct state variables' path from trajectories of Fig.~\ref{fig:DOASatUnsat}. The corrective direction at time step $k$ (Each time step duration is $T_d$) follows,

\begin{align}
\label{OmegaCorr}
 \Delta\omega_{\rm corr}(k)= \cfrac{T_d}{2H}(\Delta P_{\rm ref}(k-1))
\end{align}
\begin{align}
\label{PhaseCorrective}
\Delta\theta_{\rm corr}(k) = \Delta\theta_c(k)-\Delta\theta_c(k-1)  
\end{align}
where $\Delta P_{\rm ref}(k)$ and $\Delta\theta_c(k)$ are reference changes and corrective phase jump.

\section{MPC for Transient Stability Enhancement}
\label{sec:MPC_TS_Enhancement}

In this section, the formulation of an MPC approach to appropriately modify post-fault system trajectories is presented.

Notation-wise, $k \in \left\lbrace 0, 1, \ldots, {T}/{T_d}-1 \right\rbrace$ is used as the time index within the MPC horizon ($T$ thus being the absolute length of the rolling horizon, and $T_d$ being the time step of such a discretized horizon). $\kappa$ is used as the absolute time index, even if we often keep this index as tacit (in other words, every instant $k$ within the MPC rolling time window refers to the absolute time $\kappa + k$).

The MPC controller will thus provide for every $\kappa$ two reference vectors $\Delta P_{\rm ref}(k)$ and $\Delta \theta_c(k)$, $k \in \left\lbrace 0, 1, \ldots, \frac{T}{T_d}-1 \right\rbrace$, built to minimize the sum of the squares of the APC angle deviation from the post-fault equilibrium point, i.e.,
\begin{align}
    \min_{\mathcal{D}}
    \sum_{k=0}^{{T}/{T_d}-1}
    \left(
        \theta(k)
        -
        \arcsin \left( \frac{P_0X}{V_gV_d^{\rm ref}} \right)
    \right)^2
\label{Objective Function}
\end{align}
subject to a set of constraints $\mathcal{C}$. Here the set of decision variable $\mathcal{D}$ comprises the corrective signals $\Delta P_{\rm ref}$ and $\Delta \theta_c$ in each time step and variables that model dynamics of the system. As for the set of constraints $\mathcal{C}$, they either describe the dynamical behaviour of the GFM IBR in its transient (i.e., subset $\mathcal{C}_1$) or impose physical safe range limits (i.e., subset $\mathcal{C}_2$).

To list the elements of $\mathcal{C}_1$, we note that first and foremost the key variable in the transient stability analysis is the APC angle $\theta$. Accounting for effect of $\Delta\theta_c$ as in equation~\eqref{PhaseCorrective} alongside discretizing equation~\eqref{thetaOmega}   leads to
\begin{align}
\label{PhaseAngleEquation}
 \nonumber\theta(k+1)-\theta(k)=& \\
    \Delta\theta_c(k+1) - \Delta\theta_c(k) &+T_d\omega_n(\omega(k)-\omega_0).
\end{align}
Similarly, $\mathcal{C}_1$ includes the discretized swing equation~\eqref{swing} adjusted to account for the active power reference change $\Delta P_{\rm ref}$ as equation~\eqref{OmegaCorr}, and thus
\begin{align}
\label{GeneralSwingEquationDiscerete}
& \omega(k+1) - \omega(k) = \\
& \quad \cfrac{T_d}{2H}
    \left(
        P_0
        +
        \Delta P_{\rm ref}(k)
        -
        P(k)
        -
        \frac{1}{D_p}
        (\omega(k)-\omega_0)
    \right) .  \nonumber
\end{align}

According to Subsection~\ref{TS_model}, the active power output of the GFM IBR depends on the APC angle $\theta$ and and its operational mode, i.e. whether being in current-saturation operation or not. To capture this state, a binary variable $n(k)$ is introduced, where $n(k) = 0$ and $n(k) = 1$ represent respectively current-saturation and normal modes at time $k$. As discussed in Subsection~\ref{subsec:CE_Current_Saturation}, the GFM IBR is in the current-saturation mode if $\theta(k)$ is more than  $\theta_{\rm sat}$. To capture the fact that $n$ would be zero or one depending on this binary comparison, thus the ancillary sufficiently large parameter $M$ is added, and with the relations is introduced as
\begin{align}
\label{SaturationCond}
 \theta_{\rm sat}-\theta(k) \leq M n(k)
\end{align}
\begin{align}
\label{UnsaturationCond}
  -M (1-n(k)) \le \theta_{\rm sat}-\theta(k)
\end{align}
so that $n(k) = 1$ only if $\theta(k)$ is less than  $\theta_{\rm sat}$.

The variable $n$ enables representing the active power in current-saturation and normal operation modes for the GFM IBR as:

\begin{align}
\label{PSaturationCond}
    -M n(k)
    \le
    P -
    \frac{
        V_g I_{sat}
        \cos \big( \theta(k)+\beta \big)
        }
        {1-XC\omega\omega_n}
    \leq
    M n(k)
\end{align}
\begin{align}
\label{PUnaturationCond}
  -M (1-n(k))
  \le
  P - \frac{V_gV_d^{\rm ref}}{X}\sin \big( \theta(k) \big)\le
  M \big( 1-n(k) \big)
\end{align}

Aforementioned equations form $\mathcal{C}_1$. The rest of equations in this section construct $\mathcal{C}_2$, which ensure variables of the system behave within desired safe range. Firstly, $\mathcal{C}_2$ should ensure that the APC angle estimation module can accurately track its variations. Since fast changes of APC angle makes it challenging to have an accurate estimation of the APC angle, the changes of $\theta$ should be constrained, i.e.,
\begin{align}
\label{eq:deltathetaminmax}
  \Delta\theta_{\rm min} \le \theta(k+1)-\theta(k) \le \Delta\theta_{\rm max} \; .
\end{align}

To avoid large frequency deviations, the following frequency limits are also introduced,
\begin{align}
\label{eq:omegaminmax}
  \omega_{\rm min} \le \omega \le \omega_{\rm max}.
\end{align}

Moreover, to ensure that the GFM IBR does not absorb active power from the grid, $\theta$ should be maintained within the range of greater than zero and less than the saturated zero crossing APC angle, i.e.,

\begin{align}
    0 \le \theta(k) \le \theta_{\rm ZC}^{\rm sat}.
\end{align}

When the GFM IBR is in normal operation mode, $\Delta P_{\rm ref}$ should be set to zero. Otherwise, APC loop will receive corrective signal from the proposed MPC. This is not desired because the corrective signal should be applied only in current-saturation mode, so,

\begin{align}
  -\Delta P_{\rm ref}^{\rm max} \big( 1-n(k) \big)
  \le
  \Delta P_{\rm ref}(k)
  \le
  \Delta P_{\rm ref}^{\rm max} \big( 1-n(k) \big) .
\end{align}

 When the GFM IBR returns to the normal operation mode, $\Delta\theta_c$ should be kept at a constant value to avoid an unnecessary corrective phase jump. This requires adding the following constraint:
\begin{align}
 -\Delta \theta_c^{\rm chg, max} \big(1-n(k) \big)
  \le
  \Delta\theta_c(k+1) - \Delta\theta_c(k)\le 0. 
\end{align}

Before introducing the following constraint, we note that it has then been added after realizing that the search space, without this constraint, was sufficiently large to significantly slow down the numerical computation of the optimal. The constraint imposes that once the GFM IBR transitions from current-saturation mode to the normal operation mode, it should stay in that state. This means imposing
\begin{align}
\label{eq:nchanges}
  n(k) \le n(k+1).
\end{align}

Finally, the MPC-based optimization program is characterized by a set of initial conditions [e.g., the APC angle and corrective control signals, $\theta(0)$, $\Delta\theta_c(0)$ and $\Delta P_{\rm ref}(0)$] which can be initialized by measurement or estimation. Accounting for computation delay of the optimization solver, which is assumed to be around one time step, the control action for the next time step is calculated by the MPC.

As soon as the MPC program is solved, the proposed control approach implements $\Delta\theta_c(1)$ and $\Delta P_{\rm ref}(1)$ as the corrective control signals, and the program will be updated and solved in a rolling manner until the GFM IBR returns to normal operation.

\section{Stability of the Proposed MPC}
\label{sec:S_P_MPC}

The proposed MPC promotes both stability and safety. More in details, stability is improved in the sense that the corrective signals modify the trajectory of the system in a way that the APC angle deviations from the SEP get reduced, and potentially become zero. In addition, safety is ensured through a set of equations $\mathcal{C}_2$ that includes operating below the $\theta_{\rm ZC}^{\rm sat}$ threshold, so to ensure that active power is not absorbed from the grid. Implementing the proposed MPC considerably increases thus the DOA. We note though that we here do not provide estimates or exact computations to assess the extent of such a DOA. Doing so for Mixed-Integer Non-linear MPCs is indeed a complex task, and in any case, DOA of any control law in the feasible space is a conservative estimation for the DOA of the proposed MPC. This is because of the fact that optimality of the result of the proposed MPC leads to smaller angle deviations from SEP compared to that of feasible control law. Neglecting constraints \eqref{eq:deltathetaminmax}, \eqref{eq:omegaminmax}, and \eqref{eq:nchanges}, which either are dependant on the preferences of the operator or used for reducing the size of the optimization search space, the following control law ($CL_0$) is always a feasible point of the proposed MPC:

\begin{align}
  \Delta P_{\text{ref}}(k) &= 
    \begin{cases}
      -\Delta P_{\text{ref}}^{\text{max}} & \text{if } n(k) = 0, \\
      0 & \text{if } n(k) = 1,
    \end{cases} \nonumber \\
  \Delta \theta_c(k) &= 0,  \quad \forall k
\end{align}

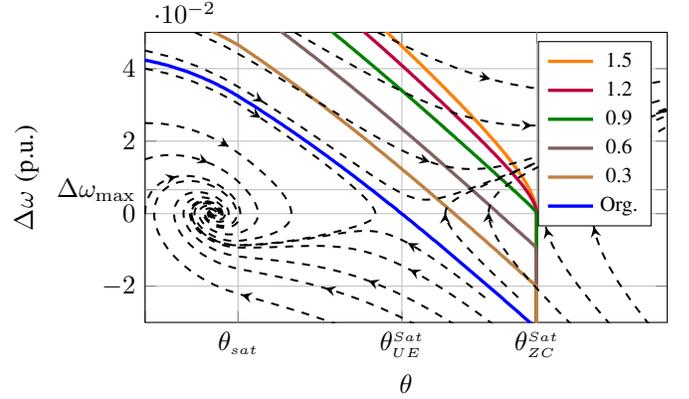
\begin{figure}[t]
    \centering
        \begin{tikzpicture}
        \begin{axis}[
            ymax = 0.05,
            ymin = -0.03,
            xmax = pi,
            xmin = 0,
            xlabel={$\theta$},
            ylabel={$\Delta\omega$ (p.u.)},
            extra y ticks={0.0066},
            extra y tick labels={$\Delta\omega_{\rm max}$},
            grid,
            legend pos=north east,
            extra x ticks={0.5593, 1.545, 2.3562},
            extra x tick labels={$\theta_{\scriptscriptstyle sat}$, $\theta_{\scriptscriptstyle UE}^{\scriptscriptstyle Sat}$, $\theta_{\scriptscriptstyle ZC}^{\scriptscriptstyle Sat}$},
            grid,
            xticklabels={},
            xtick={0}, 
            width=0.47\textwidth, height=0.3\textwidth
        ]
        
        \addplot[mark=none,line width=1.25, orange] table[col sep=space,  x expr=\thisrowno{8}, y index=9]{BodyPlot/ExtendedTrajectories/MPCConsvTraj.txt};
       \addlegendentry{\footnotesize  1.5}

        \addplot[mark=none,line width=1.25, purple] table[col sep=space,  x expr=\thisrowno{6}, y index=7]{BodyPlot/ExtendedTrajectories/MPCConsvTraj.txt};
       \addlegendentry{\footnotesize  1.2} 

        \addplot[mark=none,line width=1.25, green!50!black] table[col sep=space,  x expr=\thisrowno{4}, y index=5]{BodyPlot/ExtendedTrajectories/MPCConsvTraj.txt};
       \addlegendentry{\footnotesize  0.9} 

       \addplot[mark=none,line width=1.25, pink!50!black] table[col sep=space,  x expr=\thisrowno{2}, y index=3]{BodyPlot/ExtendedTrajectories/MPCConsvTraj.txt};
       \addlegendentry{\footnotesize  0.6}        

       \addplot[mark=none,line width=1.25, brown] table[col sep=space,  x expr=\thisrowno{0}, y index=1]{BodyPlot/ExtendedTrajectories/MPCConsvTraj.txt};
       \addlegendentry{\footnotesize  0.3} 

       \addplot[mark=none,line width=1.25, blue] table[col sep=space,  x expr=\thisrowno{0}, y index=1]{BodyPlot/ExtendedTrajectories/ExtendedSatBoundary.txt};
       \addlegendentry{\footnotesize  Org.}

      \addplot[mark=none,line width=0.75, black , dashed , postaction={ decorate, decoration={ markings, mark=at position 0.25 with {\arrow{stealth}}}}] table[col sep=space,  x expr=\thisrowno{0}, y index=1]{BodyPlot/Trajectories/SatTrajectories.txt};

      \addplot[mark=none,line width=0.75, black , dashed , postaction={ decorate, decoration={ markings, mark=at position 0.25 with {\arrow{stealth}}}}] table[col sep=space,  x expr=\thisrowno{6}, y index=7]{BodyPlot/Trajectories/SatTrajectories.txt};

      \addplot[mark=none,line width=0.75, black , dashed , postaction={ decorate, decoration={ markings, mark=at position 0.25 with {\arrow{stealth}}}}] table[col sep=space,  x expr=\thisrowno{10}, y index=11]{BodyPlot/Trajectories/SatTrajectories.txt};

     \addplot[mark=none,line width=0.75, black , dashed , postaction={ decorate, decoration={ markings, mark=at position 0.25 with {\arrow{stealth}}}}] table[col sep=space,  x expr=\thisrowno{16}, y index=17]{BodyPlot/Trajectories/SatTrajectories.txt};

     \addplot[mark=none,line width=0.75, black , dashed , postaction={ decorate, decoration={ markings, mark=at position 0.1 with {\arrow{stealth}}}}] table[col sep=space,  x expr=\thisrowno{18}, y index=19]{BodyPlot/Trajectories/SatTrajectories.txt};

    \addplot[mark=none,line width=0.75, black , dashed , postaction={ decorate, decoration={ markings, mark=at position 0.2 with {\arrow{stealth}}}}] table[col sep=space,  x expr=\thisrowno{28}, y index=29]{BodyPlot/Trajectories/SatTrajectories.txt};

    \addplot[mark=none,line width=0.75, black , dashed , postaction={ decorate, decoration={ markings, mark=at position 0.2 with {\arrow{stealth}}}}] table[col sep=space,  x expr=\thisrowno{40}, y index=41]{BodyPlot/Trajectories/SatTrajectories.txt}; 

    \addplot[mark=none,line width=0.75, black , dashed , postaction={ decorate, decoration={ markings, mark=at position 0.1 with {\arrow{stealth}}}}] table[col sep=space,  x expr=\thisrowno{50}, y index=51]{BodyPlot/Trajectories/SatTrajectories.txt};

    \addplot[mark=none,line width=0.75, black , dashed , postaction={ decorate, decoration={ markings, mark=at position 0.25 with {\arrow{stealth}}}}] table[col sep=space,  x expr=\thisrowno{6}, y index=7]{BodyPlot/ExtendedTrajectories/ExtraTrajectories.txt};

    \addplot[mark=none,line width=0.75, black , dashed , postaction={ decorate, decoration={ markings, mark=at position 0.1 with {\arrow{stealth}}}}] table[col sep=space,  x expr=\thisrowno{14}, y index=15]{BodyPlot/ExtendedTrajectories/ExtraTrajectories.txt};

        \addplot[mark=none,line width=0.75, black , dashed , postaction={ decorate, decoration={ markings, mark=at position 0.25 with {\arrow{stealth}}}}] table[col sep=space,  x expr=\thisrowno{16}, y index=17]{BodyPlot/ExtendedTrajectories/ExtraTrajectories.txt};

        \addplot[mark=none,line width=0.75, black , dashed , postaction={ decorate, decoration={ markings, mark=at position 0.25 with {\arrow{stealth}}}}] table[col sep=space,  x expr=\thisrowno{18}, y index=19]{BodyPlot/ExtendedTrajectories/ExtraTrajectories.txt};

        \addplot[mark=none,line width=0.75, black , dashed , postaction={ decorate, decoration={ markings, mark=at position 0.25 with {\arrow{stealth}}}}] table[col sep=space,  x expr=\thisrowno{20}, y index=21]{BodyPlot/ExtendedTrajectories/ExtraTrajectories.txt};
        
        \addplot[mark=none,line width=0.75, black , dashed , postaction={ decorate, decoration={ markings, mark=at position 0.25 with {\arrow{stealth}}}}] table[col sep=space,  x expr=\thisrowno{24}, y index=25]{BodyPlot/ExtendedTrajectories/ExtraTrajectories.txt};

        \addplot[mark=none,line width=0.75, black , dashed , postaction={ decorate, decoration={ markings, mark=at position 0.12 with {\arrow{stealth}}}}] table[col sep=space,  x expr=\thisrowno{26}, y index=27]{BodyPlot/ExtendedTrajectories/ExtraTrajectories.txt};

        \addplot[mark=none,line width=0.75, black , dashed , postaction={ decorate, decoration={ markings, mark=at position 0.1 with {\arrow{stealth}}}}] table[col sep=space,  x expr=\thisrowno{28}, y index=29]{BodyPlot/ExtendedTrajectories/ExtraTrajectories.txt};

         \addplot[mark=none,line width=0.75, black , dashed , postaction={ decorate, decoration={ markings, mark=at position 0.1 with {\arrow{stealth}}}}] table[col sep=space,  x expr=\thisrowno{30}, y index=31]{BodyPlot/ExtendedTrajectories/ExtraTrajectories.txt};

    \end{axis}
    \end{tikzpicture}
    \caption{The expanded boundaries of DOA provided by $CL_0$ with different per-unit values of $\Delta P_{\text{ref}}^{\text{max}}$. Black dashed lines are post-fault trajectories.}
    \label{fig:DOA_MPC}
\end{figure}

This control law though is not endowed with the possibility of using corrective phase jumps. The DOA of $CL_0$, shown in Fig.~\ref{fig:DOA_MPC}, shall be regarded as a conservative estimation of the DOA obtainable via the proposed method. In this figure the boundaries are numerically calculated for the case $X = 0.46$ p.u. This figure shows that if the post-fault frequency of the GFM IBR is kept small by bounding the frequency during the fault with a sufficiently large $\Delta P_{\rm ref}^{\rm max}$, then $CL_0$ increases the critical clearing angle up to almost $\theta_{\rm ZC}^{\rm sat}$. Therefore, the proposed MPC ensures a safely stable solution up to this post-fault APC state.

\section{Case Study}
\label{Case_Studies}

\begin{table}[!b]
\begin{center}
\caption{Parameters describing the simulated GFM farm.}
\label{GFMbasedParameters}
\begin{tabular}{ llll}
\emph{Param.} & \emph{Value}  & \emph{Unit} &  \emph{Description} \\
\hline
$S_b$   & 310 & MVA & Nominal apparent power  \\
$V_b$ &  400    &  V &  Nominal Voltage\\
$V_{dc}$  &   1200   & V  &    DC link voltage   \\
$P_0$   & 0.871    & p.u.  &  Reference active power \\
$Q_0$   &  0.0645   & p.u.  &  Reference reactive power \\
n       & 816 & - & Number of parallel GFM IBRs  \\
$f_n$   & 60 & Hz & Nominal frequency  \\
$D_p$   & 0.03    &  p.u. & Active droop coefficient  \\
$H$       & 2    &  p.u. &   Virtual inertia  \\
$D_q$ & 0.1    &  p.u. &   Reactive droop coefficient  \\
$V_0$ & 1.01     &  p.u. &   Set point voltage of RPC \\
$I_s^{max}$ & 1.2     &  p.u. &   Maximum allowed current  \\
$\beta$ & $-\cfrac{\pi}{4}$     &  Rad &  Angle of the saturated current \\
$X_{\rm tr}$ & 0.16     &  p.u. &   Total reactance of transformer windings 
\end{tabular}
\end{center}
\end{table}

To assess the effectiveness of the proposed MPC approach and mode oscillations mitigation method, simulations were conducted using Simulink/MATLAB and compared against established benchmarks. The Artelys Knitro solver was utilized to solve the mixed-integer non-linear optimization program of section~\ref{sec:MPC_TS_Enhancement}. The optimization time step is 20 ms for all cases. The simulated system models a farm of parallel, identical generating facilities (GFM IBRs) connected to a 132 kV power system via a transformer, and modeled as an equivalent GFM IBR~\cite{Tayyebi2020}. The control structure of the simulated system is the same as Fig.~\ref{fig:GridForming}. The detailed specification of the base case is summarized in Table~\ref{GFMbasedParameters}.

\subsection{Effectiveness of MPC in Enhancing Transient Stability}\label{Analysis_MPC}

This Subsection investigates to what level the proposed approach enhances the transient stability by comparing the time domain responses of the MPC added system versus the system with original controller (i.e., where no transient stability enhancement measure is employed). Since a GFM IBR can be connected to either a strong or weak grid, the transient stability enhancement that the proposed approach brings is studied for both scenarios. In this section, a strong grid and a weak grid are characterized via $Z_g = 0.3$ p.u. and $Z_g = 0.9$ p.u.; respectively. Furthermore, the proposed MPC is compared with two benchmark methods, namely, the frequency bounding approach (Strategy B) and the compensating method for the difference between saturated and unsaturated power (Strategy C), as described in Subsection~\ref{benchmark}. This is followed by sensitivity analyses and robustness analyses. Then, impact of different MPC optimization horizons is studied.

\subsubsection{Tests Under Strong Grid Conditions}\label{subsubsec:TUSGC}

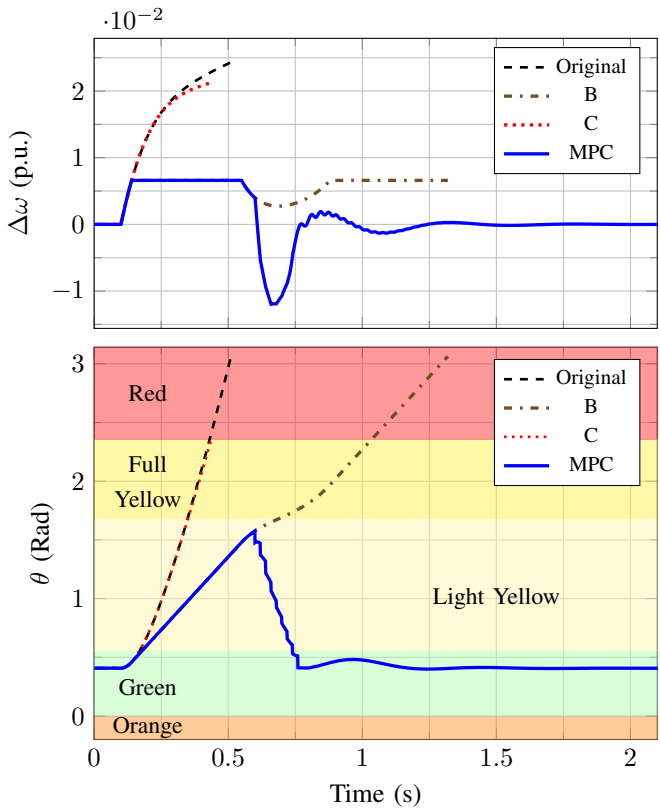
\begin{figure}[!b]
    \centering
    \begin{tikzpicture}

\begin{groupplot}[ 
    group style={
        group size=1 by 2,
        vertical sep=0.25 cm
    },
    grid=major,
    legend pos=north east,
    xmin=0,
    xmax=2.1
]

\nextgroupplot[ylabel={$\Delta\omega$ (p.u.)}, yminorgrids, xminorgrids,  xticklabels={} , minor tick num=1, width=0.5\textwidth, height=0.300\textwidth]

\addplot[mark=none,line width=1, dashed] table[col sep=space, x expr=\thisrowno{0}-3.9, y expr=\thisrowno{2}-1] {ResultPlots/Strong0_45/NoEnhancement0_45.txt};
\addlegendentry{\footnotesize Original}

\addplot[mark=none,line width=1.25, brown!60!black , dash pattern=on 3pt off 3pt on 1pt off 3pt] table[col sep=space, x expr=\thisrowno{0}-3.9,  y expr=\thisrowno{2}-1] {ResultPlots/Strong0_45/FreqBound0_45.txt};
\addlegendentry{\footnotesize B}

\addplot[mark=none,line width=1.25, red, dotted] table[col sep=space, x expr=\thisrowno{0}-3.9, y expr=\thisrowno{2}-1] {ResultPlots/Strong0_45/onlyCompensation0_45.txt};
\addlegendentry{\footnotesize C }

\addplot[mark=none,line width=1.25, blue] table[col sep=space, x expr=\thisrowno{0}-3.9, y expr=\thisrowno{2}-1] {ResultPlots/Strong0_45/Optimal0_02Timestep_0_45.txt};
\addlegendentry{\footnotesize MPC}

\nextgroupplot[ylabel={$\theta$ (Rad)}, yminorgrids, xminorgrids,  minor tick num=1, ymax=3.1415, ymin=-0.2,
xlabel={Time (s)}, width=0.5\textwidth,
  height=0.375\textwidth]

\addplot[mark=none,line width=1, dashed] table[col sep=space, x expr=\thisrowno{0}-3.9, y index=1] {ResultPlots/Strong0_45/NoEnhancement0_45.txt};
\addlegendentry{\footnotesize Original}

\addplot[mark=none,line width=1.25 , brown!60!black,  dash pattern=on 3pt off 3pt on 1pt off 3pt] table[col sep=space, x expr=\thisrowno{0}-3.9, y index=1] {ResultPlots/Strong0_45/FreqBound0_45.txt};
\addlegendentry{\footnotesize B}

\addplot[mark=none,line width=1, red, dotted] table[col sep=space, x expr=\thisrowno{0}-3.9, y index=1] {ResultPlots/Strong0_45/onlyCompensation0_45.txt};
\addlegendentry{\footnotesize C }

\path[name path=lower, opacity=0] (axis cs:\pgfkeysvalueof{/pgfplots/xmin},0) -- (axis cs:\pgfkeysvalueof{/pgfplots/xmax},0);
\path[name path=upper, opacity=0] (axis cs:\pgfkeysvalueof{/pgfplots/xmin},0.559) -- (axis cs:\pgfkeysvalueof{/pgfplots/xmax},0.559);
\addplot[green!30,opacity=0.5, forget plot] fill between[of=lower and upper];

\path[name path=lower, opacity=0] (axis cs:\pgfkeysvalueof{/pgfplots/xmin},0.559) -- (axis cs:\pgfkeysvalueof{/pgfplots/xmax},0.559);
\path[name path=upper, opacity=0] (axis cs:\pgfkeysvalueof{/pgfplots/xmin},1.6758) -- (axis cs:\pgfkeysvalueof{/pgfplots/xmax},1.6758);
\addplot[yellow!40,opacity=0.5, forget plot] fill between[of=lower and upper];

\path[name path=lower, opacity=0] (axis cs:\pgfkeysvalueof{/pgfplots/xmin},1.6758) -- (axis cs:\pgfkeysvalueof{/pgfplots/xmax},1.6758);
\path[name path=upper, opacity=0] (axis cs:\pgfkeysvalueof{/pgfplots/xmin},2.3562) -- (axis cs:\pgfkeysvalueof{/pgfplots/xmax},2.3562);
\addplot[yellow!80,opacity=0.5, forget plot] fill between[of=lower and upper];

\path[name path=lower, opacity=0] (axis cs:\pgfkeysvalueof{/pgfplots/xmin},2.3562) -- (axis cs:\pgfkeysvalueof{/pgfplots/xmax},2.3562);
\path[name path=upper, opacity=0] (axis cs:\pgfkeysvalueof{/pgfplots/xmin},3.1415) -- (axis cs:\pgfkeysvalueof{/pgfplots/xmax},3.1415);
\addplot[red!80,opacity=0.5, forget plot] fill between[of=lower and upper];

\path[name path=lower, opacity=0] (axis cs:\pgfkeysvalueof{/pgfplots/xmin},-0.2) -- (axis cs:\pgfkeysvalueof{/pgfplots/xmax},-0.2);
\path[name path=upper, opacity=0] (axis cs:\pgfkeysvalueof{/pgfplots/xmin},0) -- (axis cs:\pgfkeysvalueof{/pgfplots/xmax},0);
\addplot[orange!80,opacity=0.5, forget plot] fill between[of=lower and upper];

\addplot[mark=none,line width=1.25, blue] table[col sep=space, x expr=\thisrowno{0}-3.9, y index=1] {ResultPlots/Strong0_45/Optimal0_02Timestep_0_45.txt};
\addlegendentry{\footnotesize MPC}

\node at (axis cs:0.2,2.75) {\small{Red}};
\node at (axis cs:0.2,2.15) {\small{Full}};
\node at (axis cs:0.2,1.85) {\small{Yellow}};
\node at (axis cs:1.5,1) {\small{Light Yellow}};
\node at (axis cs:0.2,0.25) {\small{Green}};
\node at (axis cs:0.2,-0.1) {\small{Orange}};

\end{groupplot}

\end{tikzpicture}
    \caption{APC state variables with different control strategies for the case of connection to a strong grid.}
    \label{fig:AngleandFrequency_strong}
\end{figure}

According to Fig.~\ref{fig:AngleandFrequency_strong}, a fault occurrence at $t_0 = 0.1$ s and lasting 450 ms , which is simulated by decreasing the Thevenin's voltage to 0.05 p.u., results in an increase in the GFM IBRs' APC angle from 0.407 Rad to 1.574 Rad. In the absence of corrective measures, the APC angle surpasses its instability limit even before the fault clearance. In this case the \emph{compensating control} strategy (i.e., C) mitigates this effect by slightly decreasing the acceleration. However, the method is not efficient in long fault scenarios where the difference between on-fault saturated and unsaturated power is not significant enough to ensure stability. In contrast, for shorter faults, a considerable discrepancy between post-fault unsaturated and post-fault saturated power would assist in maintaining system stability by providing more deceleration.

In this case, setting the frequency bound (Strategy B) to 1.0066 p.u.\ limits the APC angle increment rate. However, due to insufficient post-fault deceleration, the system passes the unstable equilibrium in current-saturation mode shortly after the fault, as depicted by the transition in  Fig.~\ref{fig:AngleandFrequency_strong} from the light yellow to the full yellow area. More precisely, the area colors in this figure indicate:

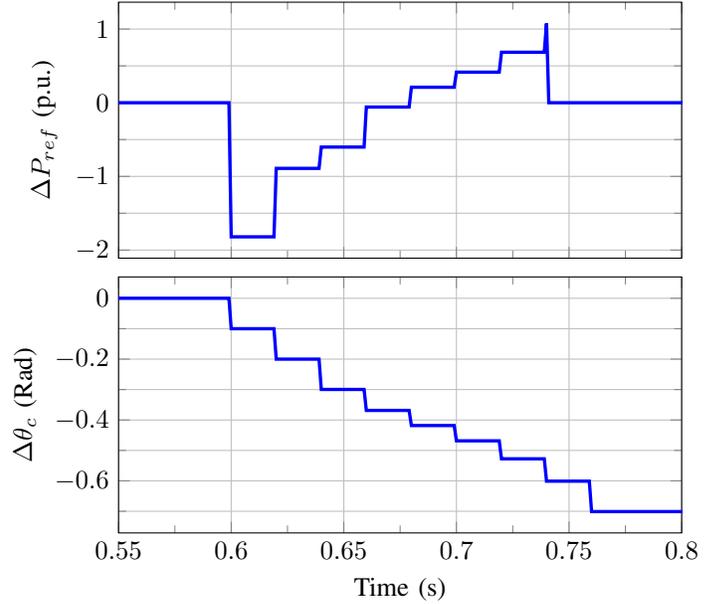
\begin{figure}[!t]
    \centering
    \begin{tikzpicture}

\begin{groupplot}[ 
    group style={
        group size=1 by 2,
        vertical sep=0.25 cm
    },
    xlabel={Time (s)},
    grid=major,
    legend pos=north east,
]

\nextgroupplot[ylabel={$\Delta P_{ref}$ (p.u.)}, xmin = 0.55, xmax = 0.8, xticklabels={} , width=0.5\textwidth,height=0.275\textwidth ,
   xlabel={}, yminorgrids, minor tick num=1]

\addplot[mark=none,line width=1.25, blue] table[col sep=space, x expr=\thisrowno{0}-3.9, y index=1]{ResultPlots/Strong0_45/deltaPref_MPC0_02Timestep_0_45.txt};

\nextgroupplot[ylabel={$\Delta \theta_c$ (Rad)}, yminorgrids, minor tick num=1, xmin = 0.55, xmax = 0.8, width=0.5\textwidth,
  height=0.275\textwidth]

\addplot[mark=none,line width=1.25, blue] table[col sep=space, x expr=\thisrowno{0}-3.9, y index=1] {ResultPlots/Strong0_45/deltaThetaC_MPC0_02Timestep_0_45.txt};

\end{groupplot}

\end{tikzpicture}
    \caption{The optimal values for the corrective control law for the strong grid-connected GFM simulation in Fig~.\ref{fig:AngleandFrequency_strong}.}
    \label{fig:CorrectiveControlLawStrong}
\end{figure}

\begin{itemize}

\item green: a GFM IBR operating normally (i.e., unsaturated) after the fault;

\item light yellow: a GFM IBR operating in the current-saturation post-fault operation. In this case, the APC angle is smaller than the saturated unstable equilibrium, there may be a possibility to maintain stability.

\item full yellow: a situation where the APC angle is greater than the unstable saturated-mode equilibrium. However, this situation is so that only with appropriate compensating control, the system may return to a normal operation mode.

\item red: a situation where the APC angle is greater than the active power crossing zero. In this area, the GFM IBR absorbs active power from the grid, which is hazardous for the DC link capacitor. Although there may be a chance to return to normal operation with corrective control, operating in this area is unstable because it is \emph{unsafe}.

\item orange: a situation where the active power is negative, and the GFM IBR absorbs power from the grid. Although this is an undesired situation, it is still an acceptable occurrence when the GFM IBR briefly enters this area and returns to the green area. The maximum duration that GFM IBR is allowed to enter this area depends on the capacitance of DC link capacitor and its maximum tolerable voltage. This situation occurs during substantial deceleration with minor damping.

\end{itemize}

The implementation of MPC effectively restores the GFM IBR to the green area, as depicted in Fig.~\ref{fig:AngleandFrequency_strong}. The corrective control law during the saturated mode operation, as presented in Fig.~\ref{fig:CorrectiveControlLawStrong}, comprises changes in reference power and corrective phase jumps. The corrective phase jump facilitates a prompt return to operating point APC angle values, while the reference power changes dampen frequency deviations as expected. Right after the fault clearance, the proposed MPC decides on a negative $\Delta P_{\rm ref}$ to provides more deceleration. After the APC angle $\theta$ declines considerably, $\Delta P_{\rm ref}$ becomes positive to prohibit under-frequency. Meanwhile, $\Delta \theta_c$ declines to bring the APC angle $\theta$ back to the SEP.

\subsubsection{Tests Under Weak Grid Conditions}

\begin{figure}[!b]
    \centering
    \begin{tikzpicture}

\begin{groupplot}[ 
    group style={
        group size=1 by 2,
        vertical sep=0.25 cm
    },
    grid=major,
    legend pos=north east,
    xmin=0,
    xmax=1.4 ]
    
\nextgroupplot[ylabel={$\Delta \omega$ (p.u.)}, yminorgrids, xminorgrids, minor tick num=1, xticklabels={} , width=0.5\textwidth, height=0.300\textwidth]

\addplot[mark=none,line width=1, dashed] table[col sep=space, x expr=\thisrowno{0}-3.9,  y expr=\thisrowno{2}-1]{ResultPlots/Weak0_250/NoEnhancement0_250Weak.txt};
\addlegendentry{\footnotesize Original}

\addplot[mark=none,line width=1.25 , brown!60!black,  dash pattern=on 3pt off 3pt on 1pt off 3pt] table[col sep=space, x expr=\thisrowno{0}-3.9,  y expr=\thisrowno{2}-1] {ResultPlots/Weak0_250/FreqBound0_250Weak.txt};
\addlegendentry{\footnotesize B}

\addplot[mark=none,line width=1, red,  dotted] table[col sep=space, x expr=\thisrowno{0}-3.9,  y expr=\thisrowno{2}-1] {ResultPlots/Weak0_250/onlyCompensating0_250Weak.txt};
\addlegendentry{\footnotesize C}

\addplot[mark=none,line width=1.25, blue] table[col sep=space, x expr=\thisrowno{0}-3.9,  y expr=\thisrowno{2}-1] {ResultPlots/Weak0_250/MPC0_250Weak0_02.txt};
\addlegendentry{\footnotesize MPC }

\nextgroupplot[ylabel={$\theta$ (Rad)}, yminorgrids, xminorgrids,  minor tick num=1, ymax=3.1415, ymin=-0.2,
xlabel={Time (s)}, width=0.5\textwidth,
  height=0.375\textwidth]

\addplot[mark=none,line width=1, dashed] table[col sep=space, x expr=\thisrowno{0}-3.9, y index=1] {ResultPlots/Weak0_250/NoEnhancement0_250Weak.txt};
\addlegendentry{\footnotesize Original}

\addplot[mark=none,line width=1.25 , brown!60!black  , dash pattern=on 3pt off 3pt on 1pt off 3pt] table[col sep=space, x expr=\thisrowno{0}-3.9, y index=1] {ResultPlots/Weak0_250/FreqBound0_250Weak.txt};
\addlegendentry{\footnotesize B}

\addplot[mark=none,line width=1, red, dotted] table[col sep=space, x expr=\thisrowno{0}-3.9, y index=1] {ResultPlots/Weak0_250/onlyCompensating0_250Weak.txt};
\addlegendentry{\footnotesize C}

\path[name path=lower, opacity=0] (axis cs:\pgfkeysvalueof{/pgfplots/xmin},0) -- (axis cs:\pgfkeysvalueof{/pgfplots/xmax},0);
\path[name path=upper, opacity=0] (axis cs:\pgfkeysvalueof{/pgfplots/xmin},1.4095) -- (axis cs:\pgfkeysvalueof{/pgfplots/xmax},1.4095);
\addplot[green!30,opacity=0.5, forget plot] fill between[of=lower and upper];

\path[name path=lower, opacity=0] (axis cs:\pgfkeysvalueof{/pgfplots/xmin},1.4095) -- (axis cs:\pgfkeysvalueof{/pgfplots/xmax},1.4095);
\path[name path=upper, opacity=0] (axis cs:\pgfkeysvalueof{/pgfplots/xmin},1.89) -- (axis cs:\pgfkeysvalueof{/pgfplots/xmax},1.89);
\addplot[yellow!40,opacity=0.5, forget plot] fill between[of=lower and upper];

\path[name path=lower, opacity=0] (axis cs:\pgfkeysvalueof{/pgfplots/xmin},1.89) -- (axis cs:\pgfkeysvalueof{/pgfplots/xmax},1.89);
\path[name path=upper, opacity=0] (axis cs:\pgfkeysvalueof{/pgfplots/xmin},2.3562) -- (axis cs:\pgfkeysvalueof{/pgfplots/xmax},2.3562);
\addplot[yellow!80,opacity=0.5, forget plot] fill between[of=lower and upper];

\path[name path=lower, opacity=0] (axis cs:\pgfkeysvalueof{/pgfplots/xmin},2.3562) -- (axis cs:\pgfkeysvalueof{/pgfplots/xmax},2.3562);
\path[name path=upper, opacity=0] (axis cs:\pgfkeysvalueof{/pgfplots/xmin},3.1415) -- (axis cs:\pgfkeysvalueof{/pgfplots/xmax},3.1415);
\addplot[red!80,opacity=0.5, forget plot] fill between[of=lower and upper];

\path[name path=lower, opacity=0] (axis cs:\pgfkeysvalueof{/pgfplots/xmin},-0.2) -- (axis cs:\pgfkeysvalueof{/pgfplots/xmax},-0.2);
\path[name path=upper, opacity=0] (axis cs:\pgfkeysvalueof{/pgfplots/xmin},0) -- (axis cs:\pgfkeysvalueof{/pgfplots/xmax},0);
\addplot[orange!80,opacity=0.5, forget plot] fill between[of=lower and upper];

\addplot[mark=none,line width=1.25, blue] table[col sep=space, x expr=\thisrowno{0}-3.9, y index=1] {ResultPlots/Weak0_250/MPC0_250Weak0_02.txt};
\addlegendentry{\footnotesize MPC }

\node at (axis cs:0.1,2.75) {\small{Red}};
\node at (axis cs:0.1,2.23) {\small{Full}};
\node at (axis cs:0.1,2.03) {\small{Yellow}};
\node at (axis cs:0.8,1.75) {\small{Light Yellow}};
\node at (axis cs:0.1,0.5) {\small{Green}};
\node at (axis cs:0.1,-0.1) {\small{Orange}};

\end{groupplot}

\end{tikzpicture}
    \caption{APC state variables with different control strategies for the case of connection to a weak grid.}
    \label{fig:AngleandFrequency_weak}
\end{figure}
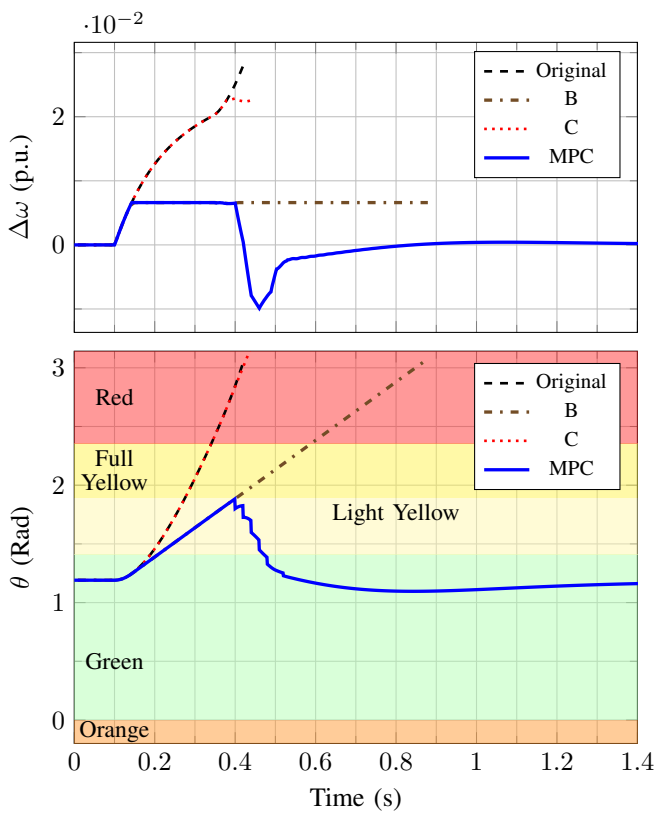

\begin{figure}[!t]
    \centering
    \begin{tikzpicture}

\begin{groupplot}[ 
    group style={
        group size=1 by 2,
        vertical sep=0.25 cm
    },
    xlabel={Time (s)},
    grid=major,
    legend pos=north east,
]

\nextgroupplot[ylabel={$\Delta P_{ref}$ (p.u.)}, xmin = 0.381, xmax = 0.55, yminorgrids, minor tick num=1, xticklabels={} ,width=0.5\textwidth,
  height=0.275\textwidth, xlabel={}]

\addplot[mark=none,line width=1.25, blue] table[col sep=space, x expr=\thisrowno{0}-3.9, y index=1]{ResultPlots/Weak0_250/deltaPrefMPC0_250Weak0_02.txt};

\nextgroupplot[ylabel={$\Delta \theta_c$ (Rad)}, yminorgrids, minor tick num=1, xmin = 0.381, xmax = 0.55, width=0.5\textwidth,
  height=0.275\textwidth]

\addplot[mark=none,line width=1.25, blue] table[col sep=space, x expr=\thisrowno{0}-3.9, y index=1] {ResultPlots/Weak0_250/deltaThetaC_MPC0_250Weak0_02.txt};

\end{groupplot}

\end{tikzpicture}
    \caption{The optimal values for the corrective control law for the weak grid-connected GFM simulation in Fig~.\ref{fig:AngleandFrequency_weak}.}
    \label{fig:CorrectiveControlLawWeak}
\end{figure}
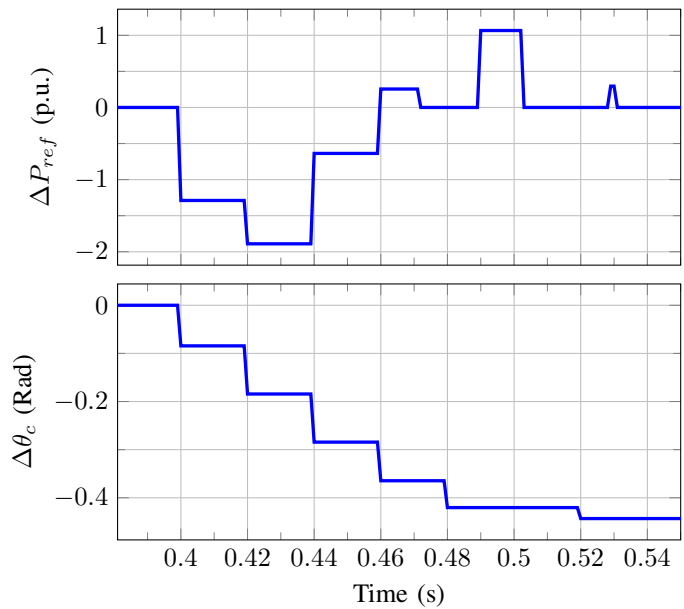

Fig.~\ref{fig:AngleandFrequency_weak} depicts the transient response of a GFM IBR connected to a weak grid when subjected to a fault that lasts 250 ms. In contrast to the strong grid situation, now the inverter's response to the fault displays a greater propensity toward instability, this is primarily because of the fact that the initial APC angle is higher and closer to the instability boundary. Consequently, even a relatively brief fault event can readily cause the inverter to cross this boundary.

During the fault event in this weak grid, the GFM IBR does not enter the current-saturation mode, as discussed in Subsection~\ref{subsec:CE_Current_Saturation}. Consequently, the fault-on behavior of strategy C remains analogous to the original strategy. However, after the fault, the GFM IBR transitions into current-saturation mode. The GFM IBR has already entered the unsafe area (red) if it is controlled through strategies original or C. The strategy B in this case limits frequency increases, but due to its insufficient post-fault deceleration properties, the system transitions from the light yellow to the full yellow area within a few milliseconds after the fault. On the other hand, the MPC corrective control, as shown in Fig.~\ref{fig:CorrectiveControlLawWeak}, offers an objective-optimal reference power change and phase angle jump during the current-saturation mode, enabling the GFM IBR to return to the normal operating point swiftly.

\subsubsection{Sensitivity Analyses}

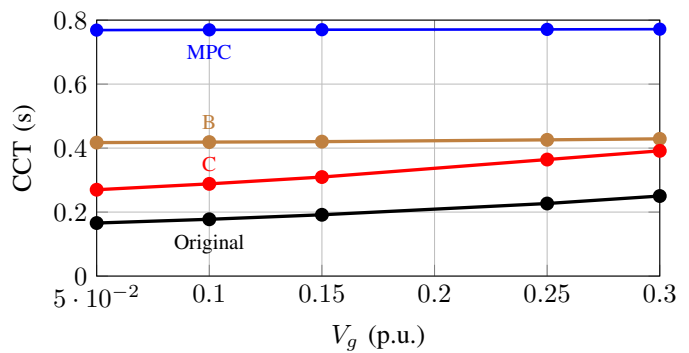
\begin{figure}[!b]
    \centering
        \begin{tikzpicture}
        \begin{axis}[
            xlabel={$V_g$ (p.u.)},
            ylabel={CCT (s)},
            xmax = 0.3 ,
            xmin = 0.05,
            ymin = 0,
            ymax = 0.8,
            grid=major , 
            width=0.5\textwidth,
            height=0.275\textwidth
        ]

        \addplot[mark=*,line width=1.25] table[col sep=space, x index=0, y index=1] {ResultPlots/CCTStrong/CCTStrong_Vg.txt};
        \node[text=black] at (axis cs:0.1,0.1) {\footnotesize Original};

        \addplot[mark=*,line width=1.25, brown] table[col sep=space, x index=0, y index=2] {ResultPlots/CCTStrong/CCTStrong_Vg.txt};
        \node[text=brown] at (axis cs:0.1,0.481) {\footnotesize B};

        \addplot[mark=*,line width=1.25, red] table[col sep=space, x index=0, y index=3] {ResultPlots/CCTStrong/CCTStrong_Vg.txt};
        \node[text=red] at (axis cs:0.1,0.35) {\footnotesize C};

        \addplot[mark=*,line width=1, blue] table[col sep=space, x index=0, y index=4] {ResultPlots/CCTStrong/CCTStrong_Vg.txt};
        \node[text=blue] at (axis cs:0.1,0.7) {\footnotesize MPC};

        \end{axis}
    \end{tikzpicture}
    \caption{Variation of the CCT with Thevenin's voltage for a GFM IBR connected to a strong grid ($Z_g$ = 0.3 p.u.).}
    \label{fig:CCTvsVoltageStrong}
\end{figure}

\begin{figure}[!t]
    \centering
        \begin{tikzpicture}
        \begin{axis}[
            xlabel={$P_0$ (p.u.)},
            ylabel={CCT (s)},
            xmax = 0.8 ,
            xmin = 0.4,
            ymax = 0.9,
            grid=major , 
            width=0.5\textwidth,
            height=0.275\textwidth
        ]

        \addplot[mark=*,line width=1.25] table[col sep=space, x index=0, y index=1] {ResultPlots/CCTDifferentPower/CCTPower.txt};
        \node[text=black] at (axis cs:0.65,0.2) {\footnotesize Original};

        \addplot[mark=*,line width=1.25, brown] table[col sep=space, x index=0, y index=2] {ResultPlots/CCTDifferentPower/CCTPower.txt};
        \node[text=brown] at (axis cs:0.7,0.58) {\footnotesize B};

        \addplot[mark=*,line width=1.25, red] table[col sep=space, x index=0, y index=3] {ResultPlots/CCTDifferentPower/CCTPower.txt};
        \node[text=red] at (axis cs:0.7,0.4) {\footnotesize C};

        \addplot[mark=*,line width=1, blue] table[col sep=space, x index=0, y index=4] {ResultPlots/CCTDifferentPower/CCTPower.txt};
        \node[text=blue] at (axis cs:0.7,0.7) {\footnotesize MPC};

        \end{axis}
    \end{tikzpicture}
    \caption{Variation of the CCT with the reference power for a GFM IBR connected to a grid with $Z_g$  = 0.4 p.u.}
    \label{fig:CCTvsPower}
\end{figure}
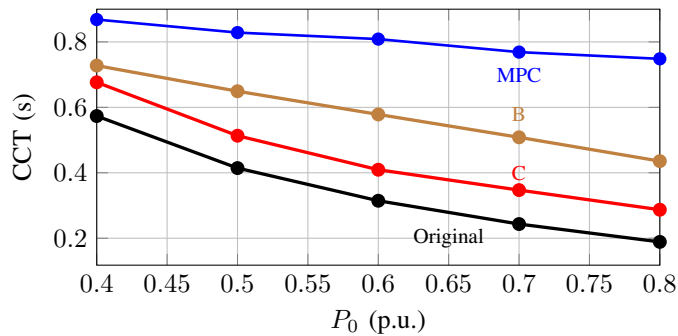

To demonstrate the capabilities of MPC in enhancing the transient stability of a GFM IBR, its performance is evaluated against the other mentioned strategies for varying fault-on Thevenin's voltages (representing the proximity of fault location in a bulk power grid to the GFM IBR).  Figs.~\ref{fig:CCTvsVoltageStrong} demonstrates a significant increase in Critical Clearing Time (CCT) due to the MPC. It shows a constant CCT around 770 ms for the GFM IBR under proposed MPC strategy. This result of simulation confirms the claim of Section~\ref{sec:S_P_MPC} about if we bound the frequency of the GFM IBR to a small value, the stability boundary is extended by the MPC up to $\theta_{\scriptsize\rm ZC}^{\rm sat}$. One can compute that it takes 783 ms of fault to increase APC angle from operating angle 0.4073 Rad to zero-crossing angle $0.75\pi$ Rad with the frequency bound of 1.0066 (p.u.). This computed time and simulated CCT are significantly close to each other. Another sensitivity analysis is also performed to ensure enhanced CCT in different loading levels (reference powers) as shown in Fig.~\ref{fig:CCTvsPower}. These two sensitivity analyses represent different operating conditions and fault locations in a bulk power system.

\subsubsection{Robustness to Parameter Estimation Error}
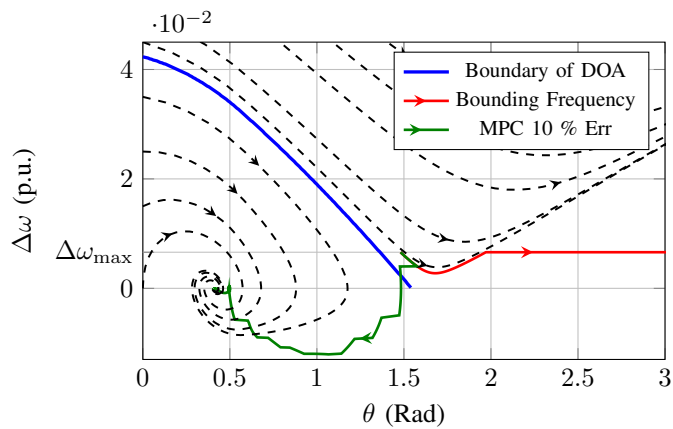
\begin{figure}[!t]
    \centering
        \begin{tikzpicture}
        \begin{axis}[
            ymax = 0.045,
            ymin = -0.013,
            xmin = 0,
            xmax = 3,
            xlabel={$\theta$ (Rad)},
            ylabel={$\Delta\omega$ (p.u.)},
            extra y ticks={0.0066},
            extra y tick labels={$\Delta\omega_{\rm max}$},
            grid,
            legend pos=north east,
            width=0.47\textwidth, height=0.32\textwidth
        ]

       \addplot[mark=none,line width=1.25, blue] table[col sep=space,  x expr=\thisrowno{0}, y index=1]{BodyPlot/Trajectories/BoundarySat.txt};
       \addlegendentry{\footnotesize Boundary of DOA}  

       \addplot[mark=none,line width=1, red , postaction={ decorate, decoration={ markings, mark=at position 0.5 with {\arrow{stealth}}}}] table[col sep=space,  x expr=\thisrowno{1}, y expr=\thisrowno{2}-1]{ResultPlots/ErrorImpedance/FreqboundControl.txt};
       \addlegendentry{\footnotesize Bounding Frequency}

      \addplot[mark=none,line width=1, green!50!black, postaction={ decorate, decoration={ markings, mark=at position 0.40 with {\arrow{stealth}}}}] table[col sep=space,  x expr=\thisrowno{1}, y expr=\thisrowno{2}-1]{ResultPlots/ErrorImpedance/OptimalControlError10percent.txt};
       \addlegendentry{\footnotesize MPC 10 \% Err}

      \addplot[mark=none,line width=0.75, black , dashed , postaction={ decorate, decoration={ markings, mark=at position 0.25 with {\arrow{stealth}}}}] table[col sep=space,  x expr=\thisrowno{0}, y index=1]{BodyPlot/Trajectories/SatTrajectories.txt};

      \addplot[mark=none,line width=0.75, black , dashed , postaction={ decorate, decoration={ markings, mark=at position 0.25 with {\arrow{stealth}}}}] table[col sep=space,  x expr=\thisrowno{6}, y index=7]{BodyPlot/Trajectories/SatTrajectories.txt};

      \addplot[mark=none,line width=0.75, black , dashed , postaction={ decorate, decoration={ markings, mark=at position 0.25 with {\arrow{stealth}}}}] table[col sep=space,  x expr=\thisrowno{10}, y index=11]{BodyPlot/Trajectories/SatTrajectories.txt};

     \addplot[mark=none,line width=0.75, black , dashed , postaction={ decorate, decoration={ markings, mark=at position 0.25 with {\arrow{stealth}}}}] table[col sep=space,  x expr=\thisrowno{14}, y index=15]{BodyPlot/Trajectories/SatTrajectories.txt};

     \addplot[mark=none,line width=0.75, black , dashed , postaction={ decorate, decoration={ markings, mark=at position 0.25 with {\arrow{stealth}}}}] table[col sep=space,  x expr=\thisrowno{18}, y index=19]{BodyPlot/Trajectories/SatTrajectories.txt};

    \addplot[mark=none,line width=0.75, black , dashed , postaction={ decorate, decoration={ markings, mark=at position 0.25 with {\arrow{stealth}}}}] table[col sep=space,  x expr=\thisrowno{20}, y index=21]{BodyPlot/Trajectories/SatTrajectories.txt};

    \addplot[mark=none,line width=0.75, black , dashed , postaction={ decorate, decoration={ markings, mark=at position 0.25 with {\arrow{stealth}}}}] table[col sep=space,  x expr=\thisrowno{34}, y index=35]{BodyPlot/Trajectories/SatTrajectories.txt}; 

    \addplot[mark=none,line width=0.75, black , dashed , postaction={ decorate, decoration={ markings, mark=at position 0.1 with {\arrow{stealth}}}}] table[col sep=space,  x expr=\thisrowno{40}, y index=41]{BodyPlot/Trajectories/SatTrajectories.txt}; 

    \addplot[mark=none,line width=0.75, black , dashed , postaction={ decorate, decoration={ markings, mark=at position 0.05 with {\arrow{stealth}}}}] table[col sep=space,  x expr=\thisrowno{46}, y index=47]{BodyPlot/Trajectories/SatTrajectories.txt};

    \end{axis}
    \end{tikzpicture}
    \caption{Post-fault APC angle for the proposed MPC with different optimization horizons.}
    \label{fig:ErrorImpedance}
\end{figure}

Except than the grid impedance, which is practically estimated online, all inputs of the proposed MPC are either measurements or parameters of the GFM IBR. Throughout the paper, it is assumed that this estimation is perfect. In this Subsection, a case in which impedance is estimated 10 \% more than its actual value is analyzed to ensure robustness. The system is identical to Subsection~\ref{subsubsec:TUSGC} except this estimation error and also a constant reference voltage instead of a reference voltage from RPC. Fig.~\ref{fig:ErrorImpedance} shows a stable post-fault APC state variables trajectory despite the impedance estimation error.

\subsubsection{Impact of Optimization Horizon Duration}

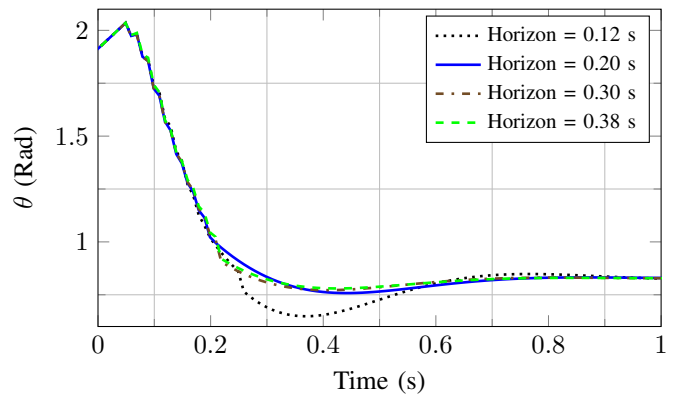
\begin{figure}[!t]
    \centering
    \begin{tikzpicture}

\begin{axis}[
ylabel={$\theta$ (Rad)}, yminorgrids, xminorgrids,  minor tick num=1, xmin=0, xmax=1, ymax=2.1, ymin=0.6,
xlabel={Time (s)}, width=0.5\textwidth, height=0.32\textwidth
        ]

\addplot[mark=none,line width=1, dotted] table[col sep=space, x expr=\thisrowno{0}-4.450, y index=1] {ResultPlots/HorizonSensitivity/OptimalHorizon0_12.txt};
\addlegendentry{\footnotesize Horizon = 0.12 s}

\addplot[mark=none,line width=1, blue] table[col sep=space, x expr=\thisrowno{0}-4.450, y index=1] {ResultPlots/HorizonSensitivity/OptimalHorizon0_20.txt};
\addlegendentry{\footnotesize Horizon = 0.20 s }

\addplot[mark=none,line width=1 , brown!60!black,  dash pattern=on 3pt off 3pt on 1pt off 3pt] table[col sep=space, x expr=\thisrowno{0}-4.450, y index=1] {ResultPlots/HorizonSensitivity/OptimalHorizon0_30.txt};
\addlegendentry{\footnotesize Horizon = 0.30 s}

\addplot[mark=none,line width=1, green, dashed] table[col sep=space, x expr=\thisrowno{0}-4.450, y index=1] {ResultPlots/HorizonSensitivity/OptimalHorizon0_38.txt};
\addlegendentry{\footnotesize Horizon = 0.38 s}

\end{axis}

\end{tikzpicture}
    \caption{Transient APC angle given different MPC optimization horizons specified in the legend.}
    \label{fig:HorizonSensitivity}
\end{figure}

In all simulations mentioned in this section, optimization horizon is 0.2 s. A large optimization horizon facilitates control decisions considering the whole trajectory. On the other hand, it causes an increase in computational burden of the required solver. To analyse the effect of this horizon, MPCs with different horizons have been performed for a case with $Z_g$ = 0.7 p.u. and fault duration of 450 ms. It is observed in Fig.~\ref{fig:HorizonSensitivity} that if the horizon becomes more than 0.2 s, the changes in the post-fault trajectory become insignificant.

\subsection{Analysis of the Capabilities in Mitigating Oscillations between Normal and Current-Saturation Operation Modes}

\begin{figure*}[!t]
    \centering
\begin{tikzpicture}

\begin{groupplot}[ 
    group style={
        group size=2 by 2,
        vertical sep=0.25 cm,
        horizontal sep=0.5 cm,
    },
    xlabel={Time (s)},
    grid=major,
    legend pos= south east,
  ]


\nextgroupplot[ylabel={$i$ (p.u.)}, xmin=0, xmax=1.6, xlabel={}, ymin = -1.3 , yminorgrids,minor tick num=1, xticklabels={} ,  width=0.5\textwidth, height=0.375\textwidth]

\addplot[mark=none,line width=1 ] table[col sep=space, x expr=\thisrowno{0}-3.9, y index=1] {ResultPlots/SaturationOscillation/CurrentOsci.txt};
\addlegendentry{\footnotesize O. $i_{d}^{ref}$}

\addplot[mark=none,line width=1 , blue] table[col sep=space, x expr=\thisrowno{0}-3.9, y index=3] {ResultPlots/SaturationOscillation/CurrentOsci.txt};
\addlegendentry{\footnotesize  O. $i_{s,d}$}

\addplot[mark=none,line width=1 , brown] table[col sep=space, x expr=\thisrowno{0}-3.9, y index=2] {ResultPlots/SaturationOscillation/CurrentOsci.txt};
\addlegendentry{\footnotesize O. $i_{q}^{ref}$}

\addplot[mark=none,line width=1 , red] table[col sep=space, x expr=\thisrowno{0}-3.9, y index=4] {ResultPlots/SaturationOscillation/CurrentOsci.txt};
\addlegendentry{\footnotesize  O. $i_{s,q}$}


\nextgroupplot[ xmin=0, yminorgrids, minor tick num=1, xmax=1.6,  ymin = -1.3 ,  xlabel={}, xticklabels={} , width=0.5\textwidth,
  height=0.375\textwidth, ylabel={},  yticklabels={}]

\addplot[mark=none,line width=1 ] table[col sep=space, x expr=\thisrowno{0}-3.9, y index=1] {ResultPlots/SaturationOscillation/CurrentNoneOsci.txt};
\addlegendentry{\footnotesize N.O. $i_{d}^{ref}$}

\addplot[mark=none,line width=1 , blue] table[col sep=space, x expr=\thisrowno{0}-3.9, y index=3] {ResultPlots/SaturationOscillation/CurrentNoneOsci.txt};
\addlegendentry{\footnotesize  N.O. $i_{s,d}$}

\addplot[mark=none,line width=1 , brown] table[col sep=space, x expr=\thisrowno{0}-3.9, y index=2] {ResultPlots/SaturationOscillation/CurrentNoneOsci.txt};
\addlegendentry{\footnotesize N.O. $i_{q}^{ref}$}

\addplot[mark=none,line width=1 , red] table[col sep=space, x expr=\thisrowno{0}-3.9, y index=4] {ResultPlots/SaturationOscillation/CurrentNoneOsci.txt};
\addlegendentry{\footnotesize  N.O. $i_{s,q}$}


\nextgroupplot[ylabel={$\theta$ (Rad)}, yminorgrids, minor tick num=1, ymax =1.2, ymin=0.3, xmin=0, xmax=1.6, width=0.5\textwidth, height=0.275\textwidth, extra description/.code={
        \node[text width=2cm, align=center] at (3.8 cm, -1.2 cm) {(a)};
    }]]

\addplot[mark=none,line width=1.25, blue] table[col sep=space, x expr=\thisrowno{0}-3.9, y index=1] {ResultPlots/SaturationOscillation/AngleOsci.txt};


\nextgroupplot[ yminorgrids, minor tick num=1, ymax =1.1, ymin=0.3, xmin=0, xmax=1.6 , width=0.5\textwidth,
  height=0.275\textwidth, ylabel={},  yticklabels={} ,   extra description/.code={
        \node[text width=2cm, align=center] at (3.8 cm, -1.2 cm) {(b)};
    }]

\addplot[mark=none,line width=1.25, blue] table[col sep=space, x expr=\thisrowno{0}-3.9, y index=1] {ResultPlots/SaturationOscillation/AngleNoneOsci.txt};

\end{groupplot}

\end{tikzpicture}
    \caption{(a) The original CRS (Oscillatory) presented \cite{Rokrok2022}, (b) The modified mode oscillations mitigation control approach (Non-Oscillatory) proposed in Subsection~\ref{subsec:Oscillations}.}
    \label{fig:SaturationOscillationRemoval}
\end{figure*}
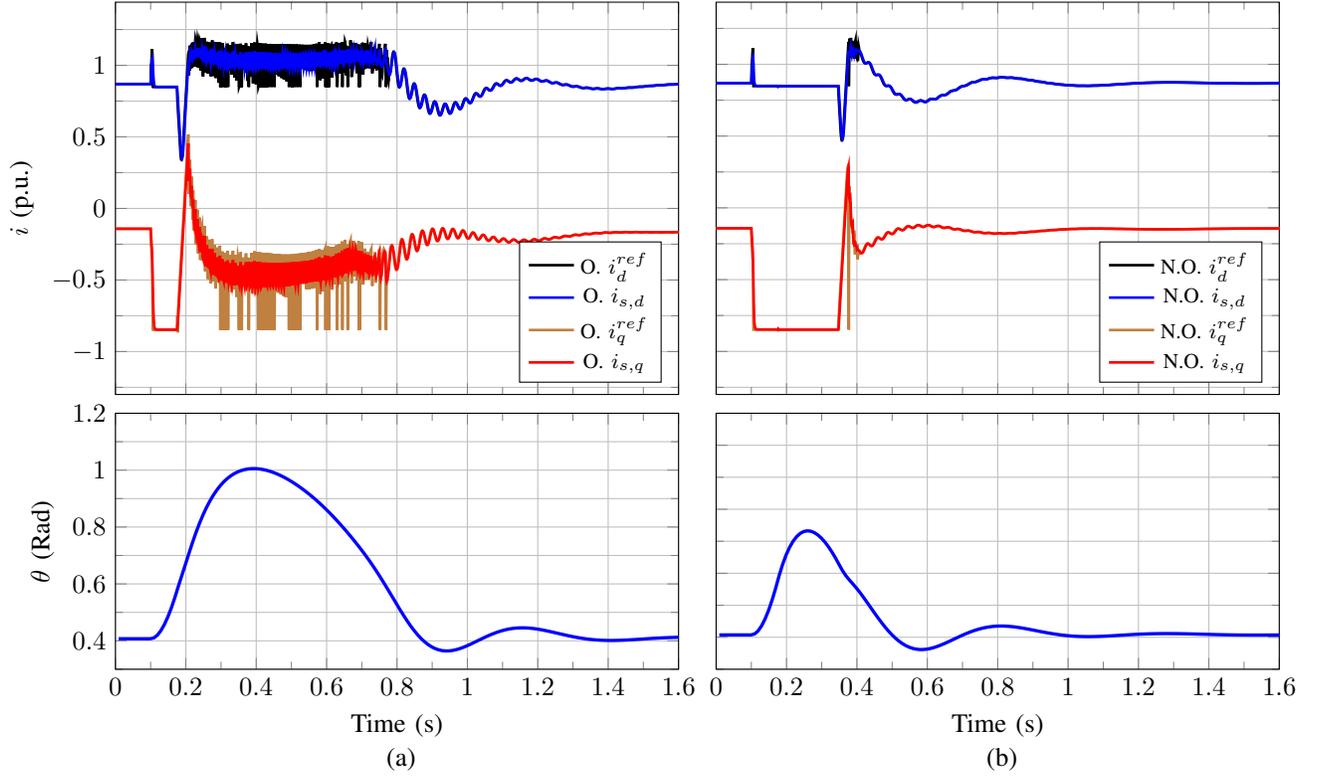

Two specific cases are analyzed in this Subsection for the purpose of evaluating the impact of the proposed mode oscillations mitigation modification of Subsection~\ref{subsec:Oscillations}. More specifically, one of the analysed cases does not incorporate the mode oscillations mitigation strategy, depicted Fig.~\ref{fig:SaturationOscillationRemoval}(a), whereas the other case does, depicted in  Fig.~\ref{fig:SaturationOscillationRemoval}(b). Both cases present the GFM IBR subjected to a three-phase fault that occurred from 0.1 s to 0.175 s. As shown by the Fig.~\ref{fig:SaturationOscillationRemoval}(a), significant oscillations is observed from 0.175 s to approximately 0.8 s. However, with the application of the proposed mode oscillations mitigation control approach, the GFM IBR remains in the current-saturation mode until reaching the condition for normal operation with just brief oscillations.

\section{Conclusions}
\label{Conclusions}

This paper presents a novel model-oriented strategy to improve the GFM IBRs transient post-fault behaviour considering current saturation, and provided a model-driven analysis of its performance. It is shown that GFM IBR's APC angle, grid voltage and grid strength are the main factors that contribute to the situation that GFM IBR's current reference exceeds its limitation. This model is used to draft an MPC approach which recursively generates an objective-optimal corrective phase jump and reference power change.

The performance of this novel scheme is analysed by means of simulations. The proposed method exhibits better performance  compared with original and benchmarks under both weak and strong grid conditions.

Although we here adopt a virtual synchronous machine approach as in~\cite{Chen2022}, the proposed methods can also be extended to other types of GFM control technologies by simple modifications in the formulation.

\section*{Acknowledgments}

Thanks Prof. Dr. Yusheng Xue for providing useful feedback on the research direction.

{\appendices
\section*{Appendix A: Definition of the Saturation Angle Threshold}

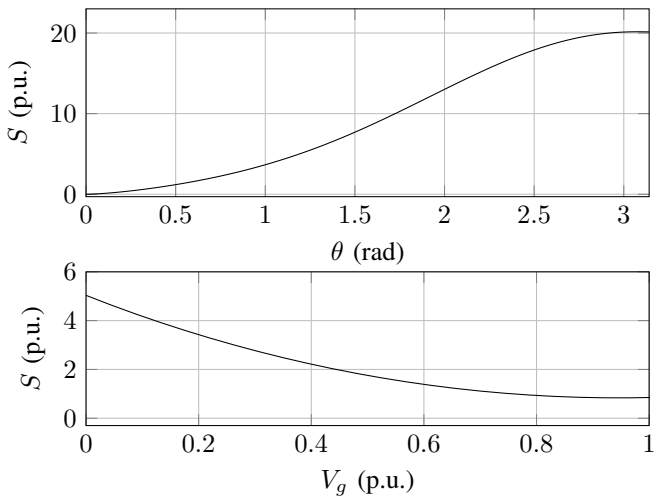
\begin{figure}[!b]
    \centering
    \scalebox{1}{\begin{tikzpicture}

    \pgfmathsetmacro{\Sb}{310000000}
    \pgfmathsetmacro{\Vg}{326.6}
    \pgfmathsetmacro{\Vref}{326.6}
    \pgfmathsetmacro{\X}{2.3742e-04} 

\begin{groupplot}[ 
    group style={
        group size=1 by 2,
        vertical sep=1 cm
    },
    grid=major,
    legend pos=north east,
]

\nextgroupplot[xlabel={$\theta$ (rad)},
        ylabel={$S$ (p.u.)},
        ymin=-0.3,
        ymax=23,
        domain=0:3.1415,
        samples=300,
        xmin = 0,
        xmax = 3.1415,
        width=0.5\textwidth,
        height=0.225\textwidth
        ]

   \addplot[domain=0:3.1415, name path = Plot1] {sqrt(1.5*(\Vg*\Vref*sin(deg(x))/\X)/\Sb)*(1.5*(\Vg*\Vref*sin(deg(x))/\X)/\Sb) + (1.5*(\Vref*\Vref-\Vg*\Vref*cos(deg(x)))/(\X*\Sb))*(1.5*(\Vref*\Vref-\Vg*\Vref*cos(deg(x)))/(\X*\Sb))};

\nextgroupplot[ylabel={$S$ (p.u.)},
        ymin=-0.3,
        ymax=6,
        xmin = 0,
        xmax = 1,
        domain=0:1.05,
        xlabel={$V_g$ (p.u.)},
        width=0.5\textwidth,
        height=0.2\textwidth
        ]

    \addplot[name path = Plot2] {sqrt(1.5*((x)*\Vg*\Vref*sin(deg(0.4))/\X)/\Sb)*(1.5*((x)*\Vg*\Vref*sin(deg(0.4))/\X)/\Sb) + (1.5*(\Vref*\Vref-(x)*\Vg*\Vref*cos(deg(0.4)))/(\X*\Sb))*(1.5*(\Vref*\Vref-(x)*\Vg*\Vref*cos(deg(0.4)))/(\X*\Sb))};

\end{groupplot}

\end{tikzpicture}}
    \caption{Variations of apparent power to voltage sag and APC angle for a GFM IBR connected to a grid ($X = 0.46$ p.u.)}
    \label{fig:PthetaQV}
\end{figure}

Initiated by a voltage sag or large APC angle, a GFM's output apparent power becomes an enormous value as indicated in Fig.~\ref{fig:PthetaQV}. In this case, the GFM IBR transitions from a normal operation mode to the current-saturation because the magnitude of the inverter side current exceeds its maximum allowed value. This current, that is computed as

\begin{align} 
\label{eq:PCC_KCL}
\mathbf{I_s} = \mathbf{I} + jC\mathbf{V}\omega_n\omega
\end{align} 
and that is formally equal to
\begin{align} 
\label{eq:PCC_KCL_comp1}
\mathbf{I_s} = \cfrac{\mathbf{V}-\mathbf{V_g}}{R+jX} + jC\mathbf{V}\omega_n\omega ,
\end{align} 
By replacing $ \mathbf{V} = V \angle \theta$ and $R+jX = Z\angle \pi$, one will reach
\begin{equation} 
\label{eq:PCC_KCL_comp2}
    \begin{array}{rcl}
        I_s &
        = &
        \phantom{+}
        \left(
              \cfrac{V}{Z}\cos{\phi}
            - \cfrac{V_g}{Z} \cos{(\theta+\phi)}
        \right)
        \\
        & &
        + j
        \left(
            CV \omega_n\omega
            + \cfrac{V_g}{Z}\sin{(\theta+\phi)}
            - \cfrac{V}{Z}\sin{\phi}
        \right) .
    \end{array}
\end{equation}

This indicates that the GFM IBR enters a current-saturation mode only if the magnitude of the inverter side current is larger than $I_s^{\mathrm{max}}$. If the grid has a high $X/R$ ratio, using equation \eqref{eq:PCC_KCL_comp2}, the saturation criterion becomes
\begin{align} 
\label{eq:sat_criteria}
    \cos{\theta}
    \leq
    \cfrac{1}{1-XC\omega_n\omega} 
    & (\frac{1}{2}( \cfrac{V_g}{V} + \cfrac{V}{V_g}) \\ 
    &  - \cfrac{ \left( ZI_s^{\mathrm{max}} \right)^2}{2V_g.V}    
        + \cfrac{VC^2\omega_n^2\omega^2Z^2}{2V_g} \nonumber \\
    &   - Z \cfrac{V}{V_g}C\omega_n\omega\sin{\phi} ). 
      \nonumber
\end{align} 
A saturation happens thus if the angle of the GFM IBR increases such that its cosine become smaller than the threshold defined by~\eqref{eq:sat_criteria}. If the effect of the filter capacitor is neglected, then the threshold becomes instead
\begin{align} 
\label{eq:sat_criteria_1}
\cos{\theta} \leq \frac{1}{2}\left(\cfrac{V_g}{V} + \cfrac{V}{V_g} \right) - \cfrac{(ZI_s^{\mathrm{max}})^2}{2V_gV} .
\end{align} 

\section*{Appendix B: Voltage and Power while in the current-saturation-Mode}

According to Kirkhoff Voltage Laws, one may in general model the behavior of the voltage at the terminal of a GFM IBR as
\begin{align}
\label{eq:PCC_KVL}
\mathbf{V} = \mathbf{V_g} + \mathbf{Z}\mathbf{I} .
\end{align} 
Combining then~\eqref{eq:PCC_KCL} and~\eqref{eq:PCC_KVL} for a grid with negligible resistance one obtains
\begin{align} 
\label{eq:PCC_Voltage}
\mathbf{V} = \cfrac{jX\mathbf{I_s} + \mathbf{V_g}}{1-XC\omega_n\omega}
\end{align} 
and
\begin{align} 
\label{eq:grid_side_current}
\mathbf{I} = \cfrac{\mathbf{I_s} - jC\omega_n\omega X \mathbf{V_g}}{1-XC\omega_n\omega} .
\end{align}

Neglecting the effect of the equivalent capacitor, the resulting voltage is computed as the vector summation of $jX\mathbf{I_s}$ and $\mathbf{V_g}$. Graphically, this may be represented as in Fig.~\ref{fig:dqSaturatedV}, where the angle between these two vectors equals to the sum of the GFM IBR's APC angle $\theta$, the current saturation angle $\beta$, and the grid equivalent impedance angle ($\phi \approx \cfrac{\pi}{2} \ rad$).  This figure depicts how the GFM IBR's voltage varies more under weak grid conditions.

The considerations above explain why,  in a transient event, the magnitude of the voltage changes as soon as the GFM IBR's APC angle changes.

If the grid is lossless, the output power is then the inner multiplication between $\mathbf{V_g}$ and $\mathbf{I}$. If the capacitance is small, $jC\omega x \mathbf{V_g}$ in~\eqref{eq:grid_side_current} can then be neglected. Since the angle between $\mathbf{V_g}$ and $\mathbf{I}$ is $\theta + \beta$, the output power may then be computed as
\begin{align} 
\label{eq:PCC_power}
P = \cfrac{I_sV_g \cos{(\theta+\beta)}}{1-XC\omega_n\omega} .
\end{align} 

\begin{figure}[!t]
    \centering
    \includegraphics[scale=0.65]{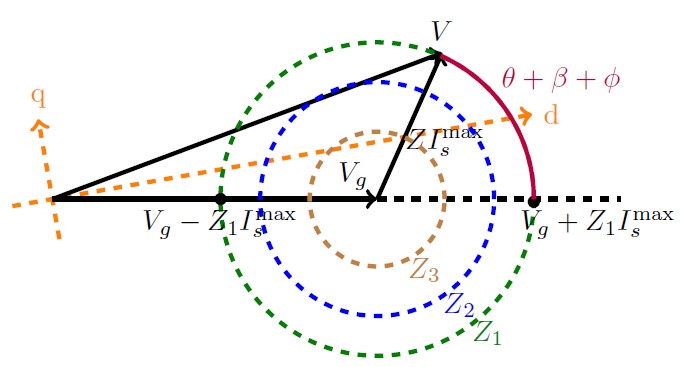}
    \caption{GFM IBR's voltage in the current-saturation mode for different grid total impedances ($Z_1 > Z_2 > Z_3$).}
    \label{fig:dqSaturatedV}
\end{figure}

}

\textbf{Ali Arjomandi-Nezhad} 
received the B.Sc. degree in electrical engineering from Amirkabir University of Technology, Tehran, Iran,
in 2016, and the M.Sc. degree in electrical engineering from Sharif University of Technology, Tehran, in 2018. In 2016, he ranked first in the Electrical Engineering Olympiad of Iran. He is currently an Early-Stage Researcher (Marie-Curie INT) and a PhD candidate at Imperial College London. His research interests include control, operation, and planning of modern power grids, especially grids with high-penetration of renewable energy resources.

\vspace{1em}

\textbf{Yifei Guo}
is a Lecturer with the School of Engineering, University of Aberdeen. He was a Postdoctoral Research Associate with the Department of Electrical and Electronic Engineering, Imperial College London, and the Department of Electrical and Computer Engineering, Iowa State University. His research interests include power system modeling, control and optimization. He is an Assistant Editor for International Journal of Electrical Power \& Energy Systems, a Young Editor for Applied Energy, and an Associate Editor for Journal of Modern Power Systems and Clean Energy.

\vspace{1em}

\textbf{Bikash Pal}
(Fellow, IEEE, Fellow RAEng) received the B.E.E. degree (Hons.) in electrical engineering from Jadavpur University, Calcutta, India, in 1990, the M.E. degree in electrical engineering from the Indian Institute of Science, Bengaluru, India, in 1992, and the Ph.D. degree in electrical engineering from Imperial College London, London, U.K., in 1999. Currently, he is a Professor with the Department of Electrical and Electronic Engineering, Imperial College London. His current research interests include renewable energy modeling and control, state estimation, and power system dynamics. He was the Vice President of Publications of the IEEE Power and Energy Society(2019-2023). He was the Editor-in-Chief of IEEE TRANSACTIONS ON SUSTAINABLE ENERGY, from 2012 to 2017, and IET Generation, Transmission and Distribution, from 2005 to 2012.

\vspace{1em}

\textbf{Damiano Varagnolo}
 received the Dr. Eng. degree in automation engineering and the Ph.D. degree in information engineering from the University of Padova respectively in 2005 and 2011. He worked as a research engineer at Tecnogamma S.p.A., Treviso, Italy during 2006-2007 and visited UC Berkeley as a scholar researcher in 2010. From March 2012 to December 2013, he worked as a post-doctoral scholar at KTH, Royal Institute of Technology, Stockholm. From January 2014 to November 2019, he worked first as Associate Senior Lecturer and then as Senior Lecturer at LTU, Luleå University of Technology in Sweden, teaching system identification and state-space based automatic control. He is now serving as Professor at NTNU in Trondheim within the Department of Engineering Cybernetics. His research interests include statistical learning, distributed optimization, and distributed nonparametric estimation, with a special focus on applications including identification and control for the built environment, learning analytics, and muscular rehabilitation.

\vfill

\end{document}